\documentclass[journal]{IEEEtran}
\IEEEoverridecommandlockouts
\usepackage{subfigure}
\usepackage{subcaption}
\usepackage{graphicx}
\usepackage{float}
\usepackage{verbatim}
\usepackage{amsmath}
\usepackage{algpseudocode}
\graphicspath{ {image/} }
\usepackage{amsthm,amsmath,amssymb,amsfonts}
\usepackage{mathrsfs}
\usepackage{algorithm}
\usepackage{color}
\usepackage{verbatim}
\usepackage{xcolor}
\usepackage{cite}
\usepackage[separate-uncertainty = true,multi-part-units = repeat, load-configurations = abbreviations]{siunitx}
\usepackage{siunitx}
\usepackage{afterpage}
\usepackage{dblfloatfix}
\usepackage{threeparttable}
\usepackage{svg}
\usepackage{lipsum}
\usepackage{booktabs}
\usepackage[colorlinks=true, linkcolor=black, citecolor=black, urlcolor=black, pdfborder={0 0 0}]{hyperref}
\usepackage{booktabs}
\usepackage{caption}
\usepackage{textcomp}
\usepackage{arydshln}
\usepackage{makecell}
\usepackage{bm}

\usepackage[explicit]{titlesec}

\ifCLASSINFOpdf
\else
\fi
\begin{document}
%
\title{Goal-Oriented Framework for Optical Flow-based Multi-User Multi-Task Video Transmission}

%
%
%

\author{\IEEEauthorblockN{Yujie~Xu, Shutong Chen, Nan Li, Yansha~Deng, Jinhong Yuan, and Robert~Schober
}
\thanks{An early version of this work was presented in part at the IEEE International Conference on Communications (ICC), June 2025 \cite{my_conf}.}
\thanks{Yujie Xu, Shutong Chen, Nan Li and Yansha Deng are with the Department of Engineering, King's College London, London WC2R 2LS, U.K. (e-mail: \{ yujie.xu, shutong.chen, nan.3.li, yansha.deng\}@kcl.ac.uk); Jinhong Yuan is with the School of Electrical Engineering and Telecommunications at University of New South Wales (e-mail: j.yuan@unsw.edu.au); Robert Schober is with the Institute for Digital Communications, FriedrichAlexander-Universität Erlangen-Nürnberg, 91058 Erlangen, Germany (e-mail:robert.schober@fau.de)(Corresponding author: Yansha Deng).}

}
\maketitle

\begin{abstract}
Efficient multi-user multi-task video transmission is an important research topic within the realm of current wireless communication systems. To reduce the transmission burden and save communication resources, we propose a goal-oriented semantic communication framework for optical flow-based multi-user multi-task video transmission (OF-GSC). At the transmitter, we design a semantic encoder that consists of a motion extractor and a patch-level optical flow-based semantic representation extractor to effectively identify and select important semantic representations. At the receiver, we design a transformer-based semantic decoder for high-quality video reconstruction and video classification tasks. To minimize the communication time, we develop a deep deterministic policy gradient (DDPG)-based bandwidth allocation algorithm for multi-user transmission. For video reconstruction tasks, our OF-GSC framework achieves a significant improvement in the received video quality, as evidenced by a 13.47\% increase in the structural similarity index measure (SSIM) score in comparison to DeepJSCC. For video classification tasks, OF-GSC achieves a Top-1 accuracy slightly surpassing the performance of VideoMAE with only 25\% required data under the same mask ratio of 0.3. For bandwidth allocation optimization, our DDPG-based algorithm reduces the maximum transmission time by 25.97\% compared with the baseline equal-bandwidth allocation scheme.
\end{abstract}


%
\IEEEpeerreviewmaketitle

\section{Introduction}
%
%
%
%
\IEEEPARstart{T}{he} escalating demand for high-quality videos, fueled by the proliferation of smart devices and visual applications such as live video streaming, video conferencing, and video surveillance, has made efficient multi-user multi-task video transmission an important task in current wireless networks \cite{overall,overall_1,overall_1_1,overall_1_2}. However, the demand for efficient and reliable multi-task video transmission has imposed considerable pressure on existing communication systems due to the tremendous data volume and high fidelity requirements of high resolution videos  \cite{overall2,overall2_1}. 

Traditional video transmission methods primarily focus on reducing data volume through various compression techniques. In the early 1990s, the adoption of JPEG for video frames gave rise to Motion JPEG (M-JPEG). To alleviate the blocking artifacts and compression inefficiency of M-JPEG, JPEG2000 was introduced in the early 2000s and subsequently applied to intra-frame video coding, leading to M-JPEG2000. More advanced codecs, such as H.264 \cite{h264} in the early 2000s and H.265 \cite{h265} in the mid 2010s, were subsequently developed to further improve compression efficiency by exploiting temporal redundancy through inter-frame prediction and motion compensation. However, these approaches remain fundamentally constrained by bit-level optimization based on Shannon’s communication framework, which limits their ability to efficiently support high-quality video transmission in stringent communication environments \cite{overall3,overall5}.

To address this challenge, goal-oriented semantic communication (GSC) has emerged as a promising method, which may lead to a revolution in wireless network design \cite{sc_1,sc_2,sc_3,sc_4,shutong_mag}. By extracting and transmitting only goal-relevant semantic representations, GSC aligns communication with task requirements while improving scalability and sustainability \cite{sc_1,sc_2,sc_3,sc_4}. Most existing GSC studies focus on single-domain, single-modality tasks, and thus have not yet addressed the multi-user, multi-task video transmission requirements in future 6G scenarios \cite{shutong_mag}.

Early GSC works for video predominantly modeled videos as sequences of independent images \cite{sc_video_img}. The DeepJSCC algorithm \cite{deepjscc} leverages Deep Neural Networks (DNNs) to directly map frames to channel symbols and demonstrate robustness in high-resolution image transmission over wireless communication channels. Nevertheless, DeepJSCC suffers significant performance degradation in low signal-to-noise ratio (SNR) scenarios due to its limited capability to dynamically adjust the transmission based on feature importance \cite{deepjscc_lm}. To address this limitation, attention mechanisms were integrated into DeepJSCC to enable dynamic emphasis on salient image regions by prioritizing transmission of perceptually important areas and suppressing less relevant content, thereby improving reconstructed image quality \cite{adjscc}. Recently, the authors in \cite{attn_video} introduced a pixel-level 3D attention map to capture spatial semantic importance across video frames. However, these approaches still rely on frame-by-frame representations, generating independent latent representations for each frame without explicitly exploiting inter-frame temporal correlations, resulting in limited transmission efficiency for video data.

To better exploit temporal correlations for efficient video transmission, VISTA \cite{gsc_video_0} was designed to only transmit background segments, behavior-related segments, and semantic location graphs that describe the relationships among detected dynamic objects, thereby incorporating temporal correlations directly. To exploit temporal correlations more precisely, a generative AI–based GSC framework \cite{gsc_video} was proposed that selects motion-aware keyframes, transmits compact semantic representation, and reconstructs the full video via semantic reconstruction and frame interpolation. However, motion modeling in this framework is primarily conducted at the pixel level, which limits its ability to capture structured and coherent temporal dynamics.

To further exploit temporal correlations and move beyond pixel-level representations, optical flow has emerged as a natural and effective approach for explicitly capturing inter-frame motion \cite{temporal_video_1,temporal_video_2}. In  \cite{temporal_video_1}, an end-to-end deep video compression framework (DVC) applies a DNN to encode optical flow between consecutive frames into a compact latent representation at the encoder and reconstructs these two consecutive frames using generated representation from encoder together with the first frame, enabling more compact representation of motion information to reduce overall bit rate while preserving reconstruction fidelity. However, this design does not account for the distortion caused by wireless channels during end-to-end training, leading to limited robustness when deployed in wireless communication system spaces. In  \cite{temporal_video_2}, deep learning enabled semantic communication systems for video transmission (DL-SC) adopted a similar optical flow-based reconstruction pipeline as DVC while introducing feature fusion and denoising modules to improve transmission quality under low SNR conditions by treating latent vectors as semantic representations. Nevertheless, existing optical flow-based approaches only modeled inter-frame redundancy and did not jointly consider intra-frame redundancy.

\begin{figure*}[h!]
    \centering
    \includegraphics[scale=0.85]{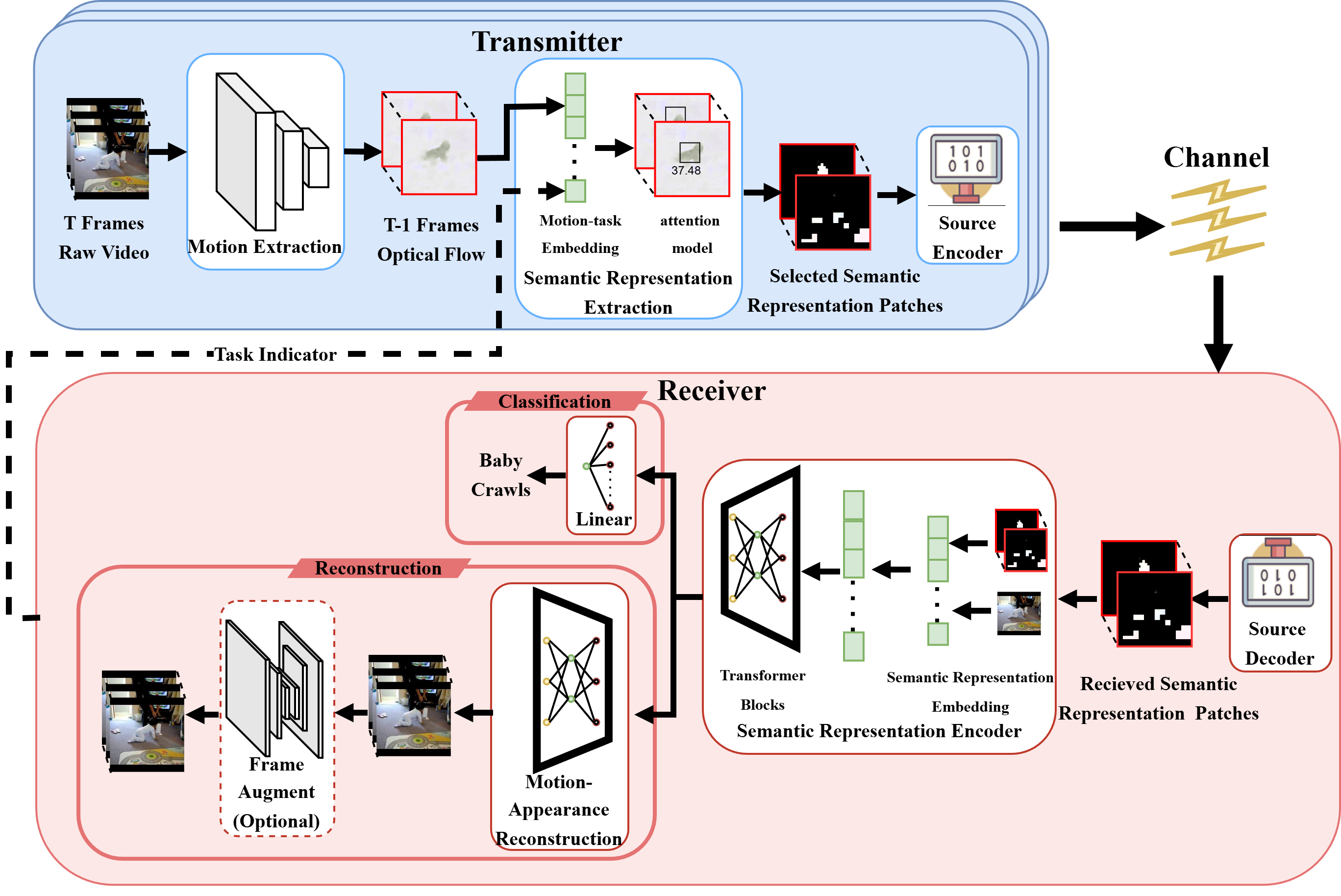}
    \centering
    \vspace{-0.2cm}
    \caption{ Optical flow-based goal-oriented semantic communication framework.}
    \label{of_gsc}
    \vspace{-0.5cm}
\end{figure*}

 In practical wireless video transmission scenarios, it is insufficient to consider the video reconstruction task alone, as the transmitted video is also expected to support other tasks, such as video classification, action recognition, and object detection. Current research on multi-task video in computer vision area primarily explores unified frameworks for jointly addressing multiple visual objectives \cite{svcm_hm,learned_svc_hm,deepsvc}. The scalable video coding for humans and machines \cite{svcm_hm} enables a single video stream to support both reconstruction and object detection by jointly considering task requirements and visual fidelity constraints in the coding process. However, this design imposes rigid structural constraints on representation allocation, which limits flexibility and coding efficiency. To improve flexibility in representation, learned scalable video coding framework \cite{learned_svc_hm} was proposed to employ jointly trained latent representations for balancing reconstruction quality and task accuracy under bit rate constraints. DeepSVC \cite{deepsvc} was further introduced, where multi-layer latent representations were employed within a shared encoding structure to support multiple visual objectives in a single model. Nevertheless, the design and training of these methods remain confined to the source coding that rely on the idealized assumptions of reliable bit-level transmission, without explicitly accounting for wireless channel impairments. As these designs focused on multi-task optimization prior to transmission, their formulation and training commonly rely on idealized or error-free channel assumptions. Consequently, the impact of wireless channel impairments, including channel noise and decoding errors, is rarely incorporated into the framework design, which limits the robustness and effectiveness of such methods in challenging wireless environments.

To fill the gap, we develop a GSC framework for optical flow-based multi-user multi-task video transmission  (OF-GSC) with the goal of achieving high-quality video reconstruction and high-accuracy video classification. Our main contributions are summarized as follows:
\begin{itemize}
\item We develop an GSC framework for optical flow-based multi-user multi-task video transmission with a semantic encoder to identify those semantic representations at the transmitter, and a semantic decoder to perform video classification or reconstruction at the receiver, respectively. To support efficient transmission of optical flow data, we design a dedicated joint encoder–decoder modular approach tailored for optical flow representation. To address the multi-user scenario, we propose a Deep Deterministic Policy Gradient (DDPG)-based bandwidth allocation algorithm that dynamically allocates bandwidth to multiple user equipments (UEs).
\item The lightweight optical flow-based semantic encoder in OF-GSC includes an optical flow-based motion extractor and a patch-level semantic representation extractor to efficiently identify important semantic representation patches with significant motion changes. We propose two alternative algorithms for semantic representation extraction, namely mathematical a motion detection algorithm and a multi-head attention method. We also design a transformer-based semantic decoder, including an optical flow-based autoencoder model to reconstruct the missing semantic representation in frames and classify the video from important semantic representation patches by using cross-model correlations and contextual information.
\item  For video reconstruction tasks, our OF-GSC framework achieves a significant improvement in generated video quality, as evidenced by a 13.47\% increase in the Structural Similarity Index Measure (SSIM) score in comparison to DeepJSCC \cite{deepjscc}. For video classification tasks, OF-GSC achieves a Top-1 accuracy that slightly surpasses the best performance of VideoMAE \cite{videomae}, while requiring only 25\% of the data used by VideoMAE under the same mask ratio of 0.3. For bandwidth allocation optimization, our DDPG-based algorithm achieves a 25.97\% reduction in the maximum transmission time compared with the baseline scheme.

\end{itemize}

The rest of this paper is organized as follows. Section II presents the framework structure for the proposed OF-GSC and the problem formulation. Section III presents the key functional modules of the OF-GSC framework. Section IV then describes the implementation of the optical-flow channel encoder and decoder, along with the strategy for channel bandwidth allocation. Section V provides simulation results and analysis. Finally, Section VI concludes the paper.

\section{System Model and Problem Formulation}
In this section, we first introduce our proposed goal-oriented semantic communication framework for optical flow-based multi-user multi-task video transmission (OF-GSC). Then, we formulate the optimization problem for efficient video reconstuction and classification.
\subsection{OF-GSC Framework}
Fig. \ref{of_gsc} illustrates our proposed OF-GSC framework for efficient wireless uplink video transmission, where we consider video reconstruction and classification tasks between multiple transmitting UEs with cameras capturing real-time videos  (e.g., 5G camera, UAV cameras) and a receiving base station (BS) over a wireless channel. At each transmitter, the semantic encoder extracts motion features from consecutive video frames, identifies goal-relevant semantic representation patches, and transmits both the compressed first frame and the coded semantic representation patches over the channel. At the receiver, the semantic decoder integrates the received data to either reconstruct the video via motion–appearance recovery or perform video classification via task-specific attention.

\subsubsection{Semantic Encoder}

The semantic encoder at the transmitter encompasses an optical flow-based motion extractor and an optical flow-based semantic representation extractor. The input video $V$ is represented as a 4D tensor $V \in \mathbb{R}^{\mathrm{T} \times \mathrm{C} \times \mathrm{H} \times \mathrm{W}}$ with $V=\{V_t\}_{t=0}^{\mathrm{T}-1}$, where $V_t$ represents the $t$-th frame of video $V$, $\mathrm{T}$ denotes the frame count, $\mathrm{C}=3$ represents the RGB color channels, and $\mathrm{H}$ and $\mathrm{W}$ are the height and width of each frame, respectively. Initially, the input video $V$ is fed into the motion extractor to extract the pixel-level motion features and output optical flow data $V_{\mathrm{OF}} \in \mathbb{R}^{(\mathrm{T}-1) \times \mathrm{C'} \times \mathrm{H} \times \mathrm{W}}$ that represent pixel-level motion information, which is given by
\begin{equation}
    V_{\mathrm{OF}}=\mathcal{E}_{\mathrm{ME}}(V),
    \label{me_general}
\end{equation}
where $\mathcal{E}_{\mathrm{ME}}(\cdot)$ denotes the motion extraction operation to extract the pixel-level motion information and $\mathrm{C'}=2<\mathrm{C}$ represents the number of motion channels in the optical flow data.

The motion-extracted data $V_{\mathrm{OF}}$ is then processed by the semantic representation extractor to identify and select the important semantic representation patches for transmission. Specifically, motion-extracted data are partitioned into a grid of non-overlapping patches with same height $\mathrm{H'}$ and same width $\mathrm{W'}$ to reduce spatial redundancy and focus on locally coherent motion patterns. The selected patch positions are denoted by $\xi \in \mathbb{R}^{\mathrm{T} \times \lceil\frac{\mathrm{H}}{\mathrm{H'}}\rceil \times \lceil\frac{\mathrm{W}}{\mathrm{W'}}\rceil}$. The semantic representation extractor masks less important patches and outputs a set of task-relevant semantic representation patches $V_{\mathrm{SR}}$ along with their spatial positions as
\begin{equation}
(V_{\mathrm{SR}},\xi)=\mathcal{E}_{\mathrm{SRE}}(V_{\mathrm{OF}}, \kappa),
\label{sif_general}
\end{equation}
where $\mathcal{E}_{\mathrm{SRE}}(\cdot)$ represents the semantic representation extraction operation and $\kappa$ denotes the task indicator specified by the receiver, which is embedded into the multi-head attention mechanism to enable task-aware feature selection. The selected semantic representation patch set $V_{\mathrm{SR}}=\{v_t\mid v_t \in \mathbb{R}^{\mathrm{C'} \times \mathrm{H'} \times \mathrm{W'}}\}$ contains task-relevant semantic representation patches for transmission, with $\mathrm{H'}\ll \mathrm{H}$ and $\mathrm{W'}\ll \mathrm{W}$ indicating that each patch has much smaller spatial dimensions than the original frames.

The first frame $V_0$, comprising the important semantic representation patch set $V_{\mathrm{SR}}$ and position information $\xi$, will be transmitted through the wireless channel to the semantic decoder at the receiver. 

\subsubsection{Wireless Channel}\label{wc}

The wireless communication channel is modeled as a Rayleigh fading channel with distance-dependent large-scale path loss. The channel coefficient $h$ is given by
\begin{equation} \label{h_mn_k}
h = \left(\frac{c}{4\pi d f_{\mathrm{c}}}\right)^{\alpha} \beta ,
\end{equation}
where $d$ denotes the distance between the transmitter and the receiver, $f_{\mathrm{c}}$ represents the uplink transmission frequency, $c$ is the speed of light, and $\alpha$ denotes the path loss exponent.
The small-scale fading coefficient $\beta$ follows a circularly symmetric complex Gaussian distribution $\mathcal{CN}(0,1)$, which characterizes Rayleigh fading.
Based on \eqref{h_mn_k}, the signal to noise ratio can be formulated as
\begin{equation}
\mathrm{SNR}=\frac{\mathrm{P} |h|^2}{\sigma^{2}},
\label{eq:SINR}
\end{equation}
where $\mathrm{P}$ denotes the transmit power and $\sigma^2$ represents the power of Additive White Gaussian Noise (AWGN).

\subsubsection{Semantic Decoder}
 The semantic decoder at the receiver includes an embedding module, a general transformer encoder, a specific transformer decoder for efficient video reconstruction, and a linear layer for video classification from the received data $(\tilde{V}_{0},\tilde{V}_\mathrm{SR},\tilde{\xi})$. 

First, the received first frame $\tilde{V}_{0}$, selected semantic representation patch set $\tilde{V}_\mathrm{SR}$, and position information $\tilde{\xi}$ are processed by the semantic representation embedding module, where the gathered  output data $V_\mathrm{EB}$ can be expressed as
\begin{equation}
V_\mathrm{EB}=\mathcal{D}_{\mathrm{EB}}(\tilde{V}_{0},\tilde{V}_\mathrm{SR},\tilde{\xi}).
\label{embedding_general}
\end{equation}
Here, $\mathcal{D}_{\mathrm{EB}}(\cdot)$ denotes the embedding module for projecting the first frame and selected semantic representation patches into patch tokens with positional embeddings for subsequent Transformer blocks.

Then, the gathered data $V_\mathrm{EB}$ is fed into the transformer blocks to analyze the spatial-temporal correlation among $(\tilde{V}_{0},\tilde{V}_\mathrm{SR},\tilde{\xi})$, where the contextual representation $V_{\mathrm{TE}}$ from transformer blocks is represented as
\begin{equation}
V_{\mathrm{TE}}=\mathcal{D}_{\mathrm{TE}}(V_\mathrm{EB}),
\label{mar_general}
\end{equation}
where $\mathcal{D}_{\mathrm{TE}}(\cdot)$ is the transformer block of the semantic representation encoder for capturing spatial temporal correlations among the embedded tokens.

Finally, the contextual representation $V_{\mathrm{TE}}$ is fed into a motion-appearance reconstruction module for video reconstruction or a linear layer processor for video classification, which generate the final output $R$ as follows
\begin{equation}
R=\left\{
\begin{array}{lr}
\mathcal{D}_{\mathrm{L}}(V_{\mathrm{TE}}), \text{Classification}\\
\\
\mathcal{D}_{\mathrm{TD}}(V_{\mathrm{TE}}),
\text{Reconstruction}
\end{array}
,
\right.
\end{equation}
where $\mathcal{D}_{\mathrm{L}}(\cdot)$ is the linear layer processor for video classification to enable lightweight task specific inference, and $\mathcal{D}_{\mathrm{TD}}(\cdot)$ is the motion-appearance module for video reconstruction to restore visual details from the shared contextual representation.
\subsubsection{Multi-task Design}
The proposed OF-GSC framework supports both video reconstruction and video classification within a unified architecture. Specifically, the semantic encoder of the transmitter and the transformer encoder of the receiver share most of the modules, including the motion extraction module, source codec and the semantic representation encoder module, to learn a common spatio-temporal representation from the received data $(\tilde{V}_{0},\tilde{V}_{\mathrm{SR}},\tilde{\xi})$. On top of these shared modules, two task-specific heads are employed, including a motion-appearance reconstruction module for video reconstruction and a lightweight linear classifier for video classification. 

To stabilize multi-task training and avoid performance interference between tasks, we adopt a stage-by-stage training strategy. The reconstruction branch is first optimized to learn general motion–appearance representations. Then, the transformer blocks of the semantic representation encoder are frozen and only the classification head with semantic representation extraction model is fine-tuned, which preserves the learned spatio-temporal features while adapting the model to the classification objective.
\subsection{Problem Formulation}
\subsubsection{Transmission Load}

To assess the transmission load across various data types, we define a transmission load metric that treats the first frame $V_0$ and the selected semantic representation patch set $V_{\mathrm{SR}}$ as numerical tensors, while using a binary tensor for the position information $\xi$. Mathematically, the transmission size $L_{\mathrm{n}}$ of $V_0$ and $V_{\mathrm{SR}}$ can be obtained as
\begin{equation}
L_{\mathrm{n}}=\underbrace{\mathrm{N}_\mathrm{b}\mathrm{H}\mathrm{W}\mathrm{C}}_{\text{Size of }V_0}+(1-\rho)\underbrace{(\mathrm{T}-1)\mathrm{N}_\mathrm{b}\mathrm{H}\mathrm{W}\mathrm{C}'}_{\text{Size of }V_{\mathrm{SR}}},
\label{trans_load1}
\end{equation}
where $\mathrm{N}_\mathrm{b}=8$ represents the bit depth of the 8-bit unsigned integer (uint8) data type used for $V_0$ and $V_{\mathrm{SR}}$; $\rho$ is the mask ratio of the semantic representation extraction module of OF-GSC. 

For binary tensor $\xi$, the transmission load requires only 1 bit per patch. Therefore, we introduce a compensation ratio $\rho_\mathrm{c}$ for binary tensor $\xi$, which is defined as
\begin{equation}
\label{comp_rat}
\rho_\mathrm{c}=\frac{1}{\mathrm{D}},
\end{equation}
where $\mathrm{D}$ is the color depth of the video source data. Therefore, transmission load $L_{\mathrm{b}}$ of binary tensor $\xi$ is
\begin{equation}
L_{\mathrm{b}}=\rho_\mathrm{c} \mathrm{N}_\mathrm{b} \mathrm{T} \frac{\mathrm{H}\mathrm{W}}{\mathrm{H}'\mathrm{W}'},
\label{trans_load3}
\end{equation}
where $L_{\mathrm{b}}$ denotes the number of bits required to convey the patch position information.

The first frame $V_0$ and the selected semantic representation patch set $V_{\mathrm{SR}}$ are coded with a compression ratio $\rho_{\mathrm{zip}}$ before transmission. Correspondingly, the transmission load $\mathrm{L}_{\mathrm{com}}$ for one video in OF-GSC can be expressed as 
\begin{equation}
\mathrm{L}_{\mathrm{com}}=(1-\rho_{\mathrm{zip}})\mathrm{L}_{\mathrm{n}}+\mathrm{L}_{\mathrm{b}}.
\label{trans_load}
\end{equation}

During transmission, the total bandwidth will be allocated for each UE based on its transmission load $\mathrm{L}_{\mathrm{com}}$ and its geographical location. The Rayleigh channel capacity is given by
\begin{equation}
\mathrm{\mathcal{C}}=\Delta t \mathrm{B}\log_{2}(1+\mathrm{SNR}),
\label{channel_cap}
\end{equation}
where $\mathrm{B}$ is the bandwidth and $\Delta t$ is the time interval for video transmission. 
\subsubsection{Video Reconstruction Task}
 In wireless video transmission, the main goal is to accurately reconstruct the original video at the receiver while minimizing distortion introduced by both OF-GSC and the wireless channel. To quantify the reconstruction quality, we use structural similarity index measure (SSIM) \cite{ssim} $f_{\mathrm{SSIM}}$ as the distortion metric
\begin{equation}
f_{\mathrm{SSIM}}(\tilde{V}_t,V_t)= \frac{(2\mu_{\tilde{V}_t} \mu_{V_t}+\mathrm{C_1})(2\sigma_{\tilde{V}_t V_t}+\mathrm{C_2})}{(\mu_{\tilde{V}_t}^2+\mu_{V_t}^2+\mathrm{C_1})(\sigma_{\tilde{V}_t}^2+\sigma_{V_t}^2+\mathrm{C_2})},
\end{equation}
where $\tilde{V}_t$ and $V_t$ are the $t$-th reconstruction frame and the $t$-th original video frame, respectively; $\mu_{\tilde{V}_t}$ and $\mu_{V_t}$ denote the mean luminance of the frames in $\tilde{V}_t$ and $V_t$; $\sigma_{\tilde{V}_t}^2$ and $\sigma_{V_t}^2$ are the variances of the frames in $\tilde{V}_t$ and $V_t$; $\sigma_{\tilde{V}_t V_t}$ denotes the covariance between $\tilde{V}_t$ and $V_t$; and $\mathrm{C_1}$ and $\mathrm{C_2}$
 are small constants to avoid division by very small numbers.

The objective of our design of wireless video transmission is to maximize the average SSIM between original and reconstructed video frames under transmission load constraints, i.e., 
\begin{equation}
\begin{aligned}
\max_{\mathcal{P}}&\frac{1}{\mathrm{T}} \sum_{i=1}^{\mathrm{T}} f_{\mathrm{SSIM}}(\tilde{V}_t,V_t),\\
&s.t.\ L_{\mathrm{com}}\leq \mathrm{L}_{\mathrm{c}},
\end{aligned}
\label{ssim}
\end{equation}
where $\mathcal{P}$ is the set of learnable parameters of the semantic decoder, and $\mathrm{L}_{\mathrm{c}}$ is the transmission load constraint.

\subsubsection{Video Classification Task}

In video classification, the main goal is to accurately classify the video at the receiver under different channel conditions. We use Top-k accuracy as the metric to quantify the classification quality. The objective is to maximize the average Top-k scores \cite{topk}, i.e.,

\begin{equation}
\begin{aligned}
\max_{\mathcal{P}}&\frac{1}{\mathrm{N}}\sum_{V \in \mathcal{V}} f_{\mathrm{Top-k}}(R_{\mathrm{C}}),\\
&s.t.\ L_{\mathrm{com}}\leq \mathrm{L}_{\mathrm{c}},
\end{aligned}
\label{ssim}
\end{equation}
where $\mathcal{V}$ is the video set, $\mathrm{N}$ is the number of videos in set $\mathcal{V}$, and $R_{\mathrm{C}}$ is the result of classification from  $\mathcal{D}_{\mathrm{L}}(\cdot)$.

\section{Optical Flow-Based Goal-Oriented Semantic Communication Framework}

In this section, we introduce the design of our OF-GSC framework in detail, which includes the motion extractor for extracting pixel-level motion features, the semantic representation extractor for identifying and selecting important semantic representation patches at the transmitter, the embedding module for projecting the received frames and patches into spatio-temporal embeddings, the motion-based goal realization module for reconstructing videos or performing task-specific inference based on motion-aware semantics, and an additional video quality augmentation module for enhancing the visual quality of the reconstructed video at the receiver.

\subsection{Motion Extraction}
\label {of_algo}
Inspired by SPyNet \cite{spynet}, we employ a lightweight coarse-to-fine network as the motion extractor to generate optical flow for pixel-level motion feature extraction from the input video $V$. The network consists of $l$ layers, where optical flow is progressively refined from the coarsest level to the finest level.

Specifically, for each time index $t$, the adjacent video frames $V_{t-1}$ and $V_t$ are successively downsampled to construct a multi-scale pyramid. At the $l$-th layer, the downsampled frames are denoted as $V_{t-1}^{l}$ and $V_t^{l}$. The optical flow estimated at the previous layer $v'_{t,l-1}$ is first upsampled using the upsampling function $u(\cdot)$ and then used to warp the frame $V_t^{l}$, yielding the warped frame $\omega(V_t^{l}, u(v'_{t,l-1}))$.

The warped frame $\omega(V_t^{l}, u(v'_{t,l-1}))$, together with the reference frame $V_{t-1}^{l}$ and the upsampled flow $u(v'_{t,l-1})$, are fed into a convolutional neural network (CNN) $G_l$ to estimate a residual flow refinement. The refined optical flow at layer $l$ is obtained by adding this residual to the upsampled flow, which can be expressed as
\begin{equation}
v'_{t,l}=G_l\big(u(v'_{t,l-1}),\omega(V_{t}^{l},u(v'_{t,l-1})),V_{t}^{l}\big)+u(v'_{t,l-1}).
\label{SPyNet}
\end{equation}

At the coarsest layer ($l=0$), the initial optical flow is directly estimated from the downsampled frame pair as
\begin{equation}
v'_{t,0}=G_0(V_{t-1}^{0},V_{t}^{0},0).
\end{equation}

By iteratively refining the optical flow across all layers, the extracted motion information is represented as
\begin{equation}
V_{\mathrm{OF}}=\{v'_{t,l} \mid 1 \le t < \mathrm{T}, \forall t \in \mathbb{N}\}.
\end{equation}

\subsection{Semantic Representation Extraction}
\label {msk_algo}
The semantic representation extractor aims to efficiently identify the important semantic representation and mask less important information at the patch-level based on the optical flow data. 
\subsubsection{Mathematic Motion Detection Algorithm}
Optical flow data indicates the magnitude and direction of the pixel-wise motion between consecutive frames. With the motion extraction result $V_\mathrm{OF}$ as input, the semantic representation extractor generates an important semantic representation patch set $V_{\mathrm{SR}}$ and the position information $\xi$ of the patches in the important semantic representation patch set. To reduce the computational load and mitigate the impact of camera movement, we design a patch-level mathematical motion detection algorithm based on the method in \cite{of_algo} to enhance stability. We assume that regions corresponding to true object motion carry more important semantic representation, while background motion mainly reflects camera movement. Accordingly, the proposed algorithm first separates foreground motion regions from the background and then selects the most informative patches within the motion regions for transmission.

For optical flow data $v'_t \in V_{\mathrm{OF}}$ from the motion extraction module, we divide optical flow data $v'_t$ into several connected but non-overlapping patches with fixed height $H'$ and width $W'$. We calculate the average optical flow value within a patch to propagate motion information from individual pixels to the entire patch as 
\begin{equation}
(p'_{\mathrm{x},i,j},p'_{\mathrm{y},i,j}) = \frac{1}{\mathrm{H}' \mathrm{W}'} \sum_{i_\mathrm{p} \in \mathrm{H}', j_\mathrm{p} \in \mathrm{W}'} (\Delta x_{i_\mathrm{p},j_\mathrm{p}},\Delta y_{i_\mathrm{p},j_\mathrm{p}}),
\label{of_ave}
\end{equation}
where $p'_{\mathrm{x},i,j}$ and $p'_{\mathrm{y},i,j}$ are the motion information of the patch; $\Delta x_{i_\mathrm{p},j_\mathrm{p}}$ and $\Delta y_{i_\mathrm{p},j_\mathrm{p}}$ are the pixel motion in the height and width directions, respectively; and $i_\mathrm{p}$ and $j_\mathrm{p}$ indicate the position of a pixel within the patch.

We consider the background optical flow to be a quadratic function of patch position, as it provides a simple but effective approximation of smooth global motion patterns induced by camera motion. Accordingly, the background optical flow is modeled as
\begin{equation}
\mathbf{p}'_{i,j} = \phi \mathbf{q},
\label{of_es}
\end{equation}
where $\textbf{p}'_{i,j}=[p'_{\mathrm{x},i,j}, p'_{\mathrm{y},i,j}]^{T}$ is the motion information of patch; $\phi$ is a $6 \times 2$ matrix that describes the optical flow field; $\textbf{q}=[i^2\ j^2\ ij\ i\ j\ 1]^{T}$  is a quadratic function vector of the patch position; and $(i,j) \in \mathbb{R}^{\lceil\frac{\mathrm{H}} {\mathrm{H'}}\rceil \times \lceil\frac{\mathrm{W}} {\mathrm{W'}}\rceil} $ is the patch position in $v'_t$.

Since both $\mathbf{p}'_{i,j}$ and $\mathbf{q}$ are available after sampling, the quadratic background motion model can be estimated via the Least Squares Regression Estimation (LSRE) method on selected patch subsets for $\mathrm{K}$ iterations. Specifically, at the $k$-th LSRE iteration, a candidate parameter matrix $\tilde{\phi}_{k}$ is obtained based on the sampled patches. These candidate estimates are further processed by the Random Sample Consensus (RANSAC) algorithm to obtain a robust background motion estimate $\phi^*$, which corresponds to the parameter matrix supported by the largest number of inlier patches and effectively suppresses the influence of foreground motion and noise.

Then, we adopt an adaptive threshold $l_{\mathrm{th}}$ to separate motion patches from background patches based on the magnitude of patch-level motion information $\textbf{p}'_{i,j}$ as follows,
\begin{equation}
l_{\mathrm{th}}=\alpha_1+\alpha_2\times \frac{1}{\mathrm{N}_\mathrm{pix}}\sqrt{p'^2_{\mathrm{x},i,j}+ p'^2_{\mathrm{y},i,j}},
\label{mag_th}
\end{equation}
where $\alpha_1$ and $\alpha_2$ are hyper-parameters determined by datasets and $\mathrm{N}_\mathrm{pix}$ denotes the number of calculated pixels.

A patch is considered an important semantic representation patch if it satisfies
\begin{equation}
|p'_{\mathrm{x},i,j}-\tilde{p}'_{\mathrm{x},i,j}|+|p'_{\mathrm{y},i,j}-\tilde{p}'_{\mathrm{y},i,j}|>l_{\mathrm{th}},
\label{m:mag}
\end{equation}
where $\tilde{p}'_{\mathrm{x},i,j}$ and $\tilde{p}'_{\mathrm{y},i,j}$ are the estimated background optical flow value using robust background motion estimate $\phi^*$ and quadratic position function $\textbf{q}$.

The magnitude metric derived in \eqref{m:mag} is generally effective but may struggle with zooming. Under camera zooming scenarios, background optical flow exhibits highly consistent directions due to global camera-induced motion and thus aligns well with the estimated background motion model $\phi^*$. In contrast, optical flow corresponding to true object motion is often different from this dominant background direction. Therefore, the angular difference between the estimated background motion and the current patch-level optical flow can be exploited to further distinguish foreground motion regions from the background. Based on this observation, an additional direction-based metric is formulated as
\begin{equation}
    \cos{(\textbf{p}',\tilde{\textbf{p}}')}<\theta_{\mathrm{th}},
\end{equation}
where $\tilde{\textbf{p}}'=[\tilde{p}'_{\mathrm{x},i,j},\tilde{p}'_{\mathrm{y},i,j}]$ is the optical flow data of the patch from \eqref{of_ave}, and $\theta_{\mathrm{th}}$ is the threshold of the angle gap with values close to 1.

Therefore, the important semantic representation patches $\textbf{p}_{\mathrm{sr}}$ are extracted as 
\begin{equation}
    \begin{aligned}
       \textbf{p}_{\mathrm{sr}}=\{\textbf{p}'_{t,i,j} | &\parallel \textbf{p}'_{t,i,j}-\tilde{\textbf{p}}'_{t,i,j}\parallel > l_{\mathrm{th}}, \\
       &\cos{(\textbf{p}'_{t,i,j},\tilde{\textbf{p}}'_{t,i,j})}<\theta_{\mathrm{th}},\\
       &\forall \textbf{p}'_{t,i,j} \in v'_t, \forall v'_t \in V_{\mathrm{OF}} \},
    \end{aligned}
    \ \ \ \ 
\label{metric}
\end{equation}
and the less important semantic representation patches $\textbf{p}_{\mathrm{lsr}}$ are extracted as
\begin{equation}
\textbf{p}_{\mathrm{lsr}} = \{\textbf{p}'_{t,i,j} | \textbf{p}'_{t,i,j} \notin \textbf{p}_{\mathrm{sr}}, \forall \textbf{p}'_{t,i,j} \in v'_t, \forall  v'_t \in V_{\mathrm{OF}} \}.
\end{equation}

After separating important semantic representation and less important semantic repersentation patches, the important semantic representation patch set $\textbf{p}_{\mathrm{sr}}$ and the less important semantic representation patch set $\textbf{p}_{\mathrm{lsr}}$ are sorted by magnitude from high to low, respectively.The important semantic representation patch set $V_{\mathrm{SR}}$ is constructed by progressively selecting patches from the important semantic representation patch set $\mathbf{p}_{\mathrm{sr}}$ and, if necessary, from the less important patch set $\mathbf{p}_{\mathrm{lsr}}$. The current mask ratio $\rho_{0}$ is updated according to the proportion of selected patches with respect to the total number of patches, and the selection process terminates once $\rho_{0}$ reaches the designated mask ratio $\rho$. The patches that are not selected are masked out, while the spatial positions of all selected patches are recorded for subsequent reconstruction. The overall algorithm is shown in \textbf{Algorithm \ref{alg:msk}}.
 
\begin{algorithm}[h]
\caption{Semantic Information Extractor}
\hspace*{0.02in}{\bf Input:}
The optical flow data $v'$
\vspace{-0.1cm}
\begin{algorithmic}[1]
\For {$t=0,...,T-1$}
    \State Calculate optical flow data $\textbf{p}_t$ of patches with \eqref{of_ave} \;
    \For {$k=0,...,\mathrm{K}$}
        \State Sample 6 patches $\textbf{p}_{\mathrm{s}}= [ \textbf{p}_{t,0} , ... , \textbf{p}_{t,5}]$ \;
        \State Get patch position data $\textbf{q}_{\mathrm{s}}$\;
        \State $\tilde{\phi}_k=\mathrm{LSRE}(\textbf{p}_{\mathrm{s}},\textbf{q}_{\mathrm{s}})$
    \EndFor
    \State $\phi^*=\mathrm{RANSAC}([\tilde{\phi}_0,...,\tilde{\phi}_{\mathrm{it}}], \textbf{p}_t)$
    \State Divide  patches $\textbf{p}_{t}$ into $\textbf{p}_{\mathrm{sr}}$ and $\textbf{p}_{\mathrm{lsr}}$
    \State Sort($\textbf{p}_{\mathrm{sr}}$, Descend)
    \State Take patches from $\textbf{p}_{\mathrm{sr}}$ until $\rho_{0}=\rho$ or all selected
    \If {$\rho_{0}<\rho$}
        \State Sort($\textbf{p}_{\mathrm{lsr}}$, Descend)
        \State Take patches from $\textbf{p}_{\mathrm{lsr}}$ until  $\rho_{0}=\rho$
    \EndIf
\EndFor
\end{algorithmic}
\label{alg:msk}
\end{algorithm}

\subsubsection{Attention-Based Model}
We propose a novel attention mechanism that extends the standard multi-head attention architecture to accommodate both classification and reconstruction goals within a unified framework.

Firstly, we design a motion-aware multi-layer perception (MLP) that explicitly models temporal relationships between frames. This motion information is directly integrated into the query computation as
\begin{equation}
\mathbf{Q}^* = \mathbf{Q}\sqrt{d_\mathrm{kv}} + \mathbf{M}'
\end{equation}
where $\mathbf{M}'$ represents the motion features processed through the MLP, and $\mathbf{Q}$ is the original query matrix with $d_\mathrm{kv}$ being the dimension of the key vectors, $\mathbf{Q}^*$ is the motion-enhanced query matrix used in our attention mechanism.

To address the dual objectives of video reconstruction and video classification, we introduce a goal-adaptive attention mechanism that dynamically adjusts the attention computation based on the specific task. The task embedding is achieved through sinusoidal positional encoding, which enables the attention mechanism to adapt its focus according to the specific objective. Based on both feature representations and the specific task objective, we introduce a unified attention model that adaptively generates patch-level semantic representation scores. The architecture comprises a task-specific adaptation layer that contextualizes features according to the current goal, a unified mask generator that produces relevance scores for each patch, and a strategic mask selection mechanism that applies a Top-k strategy on these scores. This design enables the model to selectively focus on the most task-relevant patches while efficiently managing the transmission of the semantic representation from transmitter to receiver.

\subsection{Semantic Representation Encoder}
\label{embedding}
Given the first original frame $\tilde{v}_0$, the selected important semantic representation patch set $\tilde{\textbf{p}}$ and their patch position information $\xi$, the semantic representation embedding module integrates these inputs into data $\mathcal{M}$ for the transformer blocks.

The semantic representation encoder has two components. The first component is semantic representation embedding, which integrates the first original frame $\tilde{v}_0$ and important semantic representation patch set $\tilde{\textbf{p}}$. We apply zero-padding to optical flow patch data to ensure that the motion channel $\mathrm{C}'$ matches  the color channel $\mathrm{C}$. Following patch embedding, we apply position embedding using position information $\xi$. We utilize the Cos-Sin position embedding method\cite{pos_embed} to generate distinct sine and cosine values based on position $\xi$ and then add these values to the patch-embedded data. 

The embedded representation $\mathcal{M}$ is then fed into a stack of the transformer blocks. The semantic representation encoder consists of 12 transformer blocks, where the deep encoder structure facilitates the modeling of long-range dependencies and progressively refines high-level contextual representations through successive self-attention and position-wise feed-forward operations.

\subsection{Motion-Based Goal Realization}
\label{codec}
The motion-based goal realization part is designed to support video reconstruction and video classification based on the contextual representation $V_{\mathrm{TE}}$. Specifically, the framework adopts a task-specific design, where a shallow transformer decoder is employed for motion–appearance video reconstruction, while a lightweight linear module is used for video classification.

The motion–appearance reconstruction module is composed of multiple transformer blocks, each consisting of multi-head attention and position-wise MLP layers. In OF-GSC, we employ 4 transformer blocks in the reconstruction decoder. Compared with the deep semantic representation encoder, the shallower decoder design reduces computational complexity and memory consumption, while maintaining high reconstruction quality.

\subsection{Video Quality Augmentation}

The patch-level approach for semantic representation extraction and motion–appearance reconstruction may introduce subtle discontinuities at patch boundaries due to the independent reconstruction of local regions. To address these boundary artifacts, we additionally design an optional lightweight video quality augmentation network based on U-Net \cite{unet}. The network takes the generated video and the first frame as inputs and learns to refine boundary coherence and restore structural consistency across patches. Specifically, the proposed network follows a dual-path encoder–decoder architecture with four hierarchical resolution levels. Each encoder stage progressively captures the spatial context via stacked $3\times 3$ convolutions, rectified linear unit (ReLU) activations, and stride-2 max pooling, while the decoder mirrors this process using $2\times 2$ up-convolutions followed by convolutional refinement. Skip connections between encoder and decoder layers of equal resolution preserve high-frequency details and facilitate seamless spatial integration. A final $1\times 1$ convolution projects the feature maps to the RGB color space. By leveraging multi-scale down-sampling and up-sampling operations, along with skip-connections that preserve high-resolution feature details, the proposed dual U-Net module is able to integrate the coarse motion–appearance context and fine spatial details across different scales. Since the module is applied in time-insensitive situations, it can perform deep processing and multi-scale feature fusion to correct patch boundary discontinuities and restore motion–appearance coherence.

\subsection{Loss Function}

In our framework, the training process is carefully designed with specialized loss functions for different tasks and training stages to optimize both video reconstruction and video classification performance.

For the video reconstruction task, we adopt a two-stage training strategy to balance pixel-level accuracy and perceptual quality. In first stage, we employ Mean Squared Error (MSE) loss for training, which facilitates stable convergence by minimizing the pixel-wise differences between reconstructed and original video frames. The MSE loss is defined as
\begin{equation}
\hspace{-0.1cm}
\mathcal{L}_{\mathrm{MSE}}=\frac{1}{\mathrm{THWC}} \sum_{t=1}^{\mathrm{T}} \sum_{h=1}^{\mathrm{H}} \sum_{w=1}^{\mathrm{W}} \sum_{c=1}^{\mathrm{C}}(\tilde{v}^{t,h,w,c}-v^{t,h,w,c})^2,
\end{equation}
where $\tilde{v}^{t,h,w,c}$ and $v^{t,h,w,c}$ are the reconstructed frame pixels from $\mathcal{D}_{\mathrm{TD}}(\cdot)$ and the original frame pixels, respectively.

In the second stage, we fine-tune the OF-GSC model using the SSIM as a perceptual loss. Specifically, we aim to maximize the SSIM score to improve the structural and textural quality of the reconstructed frames. Since the SSIM value is bounded above by 1, the corresponding loss function is defined as
\begin{equation}
    \mathcal{L}_{\mathrm{SSIM}}=1-f_{\mathrm{SSIM}}(\tilde{v},v).
\end{equation}
This two-stage training strategy allows the OF-GSC model to first capture broad visual content and then improve fine-grained structural details.

For the video classification task, we use the Cross-Entropy loss to train the classification head and multi-head attention model. This loss is given by
\begin{equation}
\mathcal{L}_{\mathrm{CE}}=-\sum_{n_\mathrm{c}=1}^\mathrm{N_c}G_{n_\mathrm{c}}\log(R^{\mathrm{(C)}}_{n_\mathrm{c}}),
\end{equation}
where $N_c$ is the number of video classes, $G_{n_\mathrm{c}}$ is the ground-truth label and $R^{\mathrm{(C)}}_{n_\mathrm{c}}$ is the predicted probability. 

During the classification training stage, we freeze the transformer encoder to preserve the spatiotemporal features learned during video reconstruction. This not only reduces computational cost but also prevents overfitting by leveraging generalized motion and appearance representations extracted from the semantic representation patches and optical flow.

\section{Optical Flow Codec and Bandwidth Allocation}
In this section, we introduce a joint source coding and decoding scheme tailored for optical flow data, followed by a DDPG-based bandwidth allocation algorithm.
\subsection{Optical Flow Codec}
\label{of_coder}
To efficiently transmit the important semantic representation, we introduce an innovative auto-encoder network that further captures and compresses motion information with pixel-level operation through a hierarchical structure. 

The encoder first normalizes the optical flow data to the interval $[0,1]$ by converting the optical flow vectors into polar coordinates. The normalized data is then fed into a series of convolutional blocks in the encoder. Each block consists of two convolutional layers with an interposed LeakyReLU activation function \cite{leakyrelu}. The first convolutional layer performs spatial downsampling by a factor of two and is configured with \texttt{(kernel=$3\times3$, stride=2, padding=1)}. The second convolutional layer further refines the extracted features without changing the spatial resolution and is configured with \texttt{(kernel=$3\times3$, stride=1, padding=1)}. 

Through these convolutional blocks, feature representations are extracted and aggregated to form the channel input. The LeakyReLU activation enables the network to learn nonlinear mappings from the source signal space to the coded signal space. Inspired by DeepJSCC, the output of the final convolutional layer, $V_{\mathrm{OFE}}$, is power-normalized before transmission. The normalization operation is given by
\begin{equation}
V_{\mathrm{NM}}
=
\sqrt{\gamma_{\mathrm{UE}}\,\mathrm{P}_{\mathrm{UE}}}\,
\frac{V_{\mathrm{OFE}}}
{\sqrt{V_{\mathrm{OFE}}^{\mathrm{H}} V_{\mathrm{OFE}}}},
\end{equation}
where $\gamma_{\mathrm{UE}}$ denotes a power normalization factor, $\mathrm{P}_{\mathrm{UE}}$ is the average transmit power at the user equipment, and $(\cdot)^{\mathrm{H}}$ denotes the conjugate transpose.

The decoder maps the received signal to an estimate of the original input. Similarly to the encoder, decoder blocks have one transposed convolutional layer to upsample the received data and one convolutional layer with \texttt{stride=1} to extract accurate data from noisy data.

\subsection{Bandwidth Allocation}
\label{ba}
In this subsection, we introduce the proposed Deep Reinforcement Learning (DRL)-based approach to solve the bandwidth allocation problem. We consider a multiple-input single-output (MISO) wireless communication system. Assuming that all UEs start data transmission at the same time, the primary objective of the bandwidth allocation is to transmit data as fast as possible given the available overall bandwidth. Therefore, the bandwidth allocation problem is to minimize the longest transmission time, which can be formulated as
\begin{equation}
    \begin{aligned}
    &\min_{B_1,B_2,\cdots,B_n}[\max(t_{\mathrm{tx},1},t_{\mathrm{tx},2},\cdots,t_{\mathrm{tx},n})],\\
    &\ \ \ \ \ \ s.t.\ \sum_{b=1}^{n} B_b \leq \mathrm{B}\\
    \end{aligned}
    \label{ba_problem}
\end{equation}
where $B_b$ and $t_{\mathrm{tx},b}$ are allocated bandwidth and transmission time for $b$-th UE, respectively.

We define $s \in \mathcal{S}$, $a \in \mathcal{A}$, and $r \in \mathcal{R}$ as a state, action, and reward in the proposed DRL-based algorithm, respectively, where $\mathcal{A}$ represents a continuous action space. At the beginning of the $tt$-th transmission time interval (TTI) $\left(tt\in\left\lbrace0,1,2,...\right\rbrace\right)$, the agent first observes the current state $S^{tt}$ corresponding to a set of previous observations $O^{tt}$ in order to select a deterministic action $A^{tt}\in \mathcal{A}\left(S^{tt}\right)$, where  $S^{tt}$ denotes the state representation constructed from the current observation set $O^{tt}$. Action $A^{tt}$ represents the scheduling decision $\lambda^{tt}$, which is further refined by the critic network to optimize the long-term reward.

We define the state representation for bandwidth allocation at the beginning of each TTI, where $S^{tt}$ is constructed from the observed system information $[\boldsymbol{\rho}, t_{\mathrm{max}}]$. Based on state $S^{tt}$, the agent generates a continuous control signal $A^{tt}\in\mathcal{A}$ through the actor network. 
The action $A^{tt}$ represents the bandwidth scheduling decision, expressed as a continuous allocation variable with values constrained to the interval $(0,1)$. After applying $A^{tt}$ to the environment, the agent receives a scalar reward $R^{tt+1}$ and observes the next state $S^{tt+1}$. Since the goal is to minimize the longest transmission time among all users, the reward is defined as  
\begin{equation}
R^{tt+1}=e^{-\alpha_{\mathrm{R}} t_{\mathrm{max}}},
\end{equation}
where $\alpha_{\mathrm{R}}>0$ adapts the range of transmission times and $t_{\mathrm{max}}=\max(t_1,t_2,\ldots,t_n)$ denotes the peak transmission delay among all users.

The deep deterministic policy gradient (DDPG) approach is implemented to solve the bandwidth allocation problem with continuous action space. Our DDPG algorithm consists of an actor network and a critic network. The actor network $\mu(s|\theta^\mu)$ directly maps states to action, which determines the policy by outputting a deterministic action for any given state with the parameters $\theta^\mu$. The critic network $Q(s,a|\theta^Q)$ evaluates the conducted action, which estimates the expected return of taking action $A$ with state $S$ and parameters $\theta^Q$. In our bandwidth allocation problem, the actor network outputs the bandwidth allocation proportions for each UE, which ensures the bandwidth constraint.

In deterministic policy settings, an agent that always follows the network output tends to lack sufficient exploration and may get stuck in suboptimal solutions. To encourage exploration, we introduce noise into the action selection process as
\begin{equation}
\mu'(S)=\mu(S|\theta^\mu)+\mathcal{N}_\mathrm{{DDPG}},
\end{equation}
where $\mathcal{N}_\mathrm{{DDPG}}$ is the noise for potential exploration. The noise will gradually decrease until it reaches a minimum value and will be reset in each epoch. 

After executing decided action $A^t$ in the environment, the agent observes a scalar reward $R^{t+1}$ and next state $S^{t+1}$. Each transition $(S^t, A^t, R^{t+1}, S^{t+1})$ is stored in a replay buffer \(\mathcal{M}\) of finite capacity. Once \(\lvert \mathcal{M}\rvert \ge N_{\mathrm{b}}\), we uniformly sample a minibatch of size \(N_{\mathrm{b}}\). For each sample \(p\), we construct the target as
\begin{equation}
  y^p = R^{p+1} + \gamma \,Q'\bigl(S^{p+1},\,\mu'(S^{p+1}\mid \theta^{\mu'})\mid \theta^{Q'}\bigr),
\end{equation}
where $\theta^{Q'}$ and $\theta^{\mu'}$ are the parameters of the target critic networks $Q'$ and the target actor networks $\mu'$, respectively; and $\gamma\in(0,1]$ is the discount factor accounting for future rewards. By leveraging a separate target network pair, we stabilize temporal-difference updates and mitigate oscillations in value estimation.

The critic network \(Q(S,A\mid\theta^Q)\) is then updated by minimizing the mean-squared Bellman error over the minibatch 
\begin{equation}
  L(\theta^Q) = \frac{1}{N_{\mathrm{b}}}\sum_{q=1}^{N_{\mathrm{b}}} \Bigl(y^q - Q(S^q, A^q\mid \theta^Q)\Bigr)^2,
\end{equation}
and taking a gradient step
\begin{equation}
  \theta^Q \;\leftarrow\; \theta^Q \;-\; \lambda_{\mathrm{CRI}} \,\nabla_{\theta^Q} L(\theta^Q),
\end{equation}
with critic learning rate \(\lambda_{\mathrm{CRI}}\). Concurrently, the Actor is updated via the deterministic policy gradient theorem:
\begin{equation}
  \nabla_{\theta^{\mu}}J \;\approx\;
  \frac{1}{N_{\mathrm{b}}}\sum_{q=1}^{N_{\mathrm{b}}}
    \nabla_a Q(S^q,a\mid\theta^Q)\bigl\rvert_{a=\mu(S^q)}
    \;\nabla_{\theta^{\mu}}\mu(S^q\mid\theta^{\mu}),
\end{equation}
followed by
\begin{equation}
  \theta^{\mu}\;\leftarrow\;\theta^{\mu}\;+\;\lambda_{\mathrm{ACT}}\,\nabla_{\theta^{\mu}}J,
\end{equation}
where \(\lambda_{\mathrm{ACT}}\) is the actor learning rate.

To ensure the target networks \(Q'\) and \(\mu'\) track the learned networks without abrupt changes, we apply soft updates at each learning step as
\begin{equation}
\begin{aligned}
  \theta^{Q'} &\leftarrow \tau\,\theta^Q + (1-\tau)\,\theta^{Q'}\\
  \theta^{\mu'} &\leftarrow \tau\,\theta^{\mu} + (1-\tau)\,\theta^{\mu'}
\end{aligned}
,
\end{equation}
with $(\tau\ll 1)$ controlling the interpolation between the current and target parameters. Learning proceeds over multiple episodes with the length of each episode $N_{\text{TTI}}$, until convergence in average reward or satisfaction of a predetermined performance threshold. The DDPG training algorithm is shown in \textbf{Algorithm \ref{alg:DDPG}}.

\begin{algorithm}[h]
\caption{DDPG-Based Bandwidth Allocation}
\label{alg:DDPG}
\hspace*{0.02in}{\bf Input:}
Action Range;
\begin{algorithmic}[1]
\State Algorithm hyperparameters: actor learning rate $\lambda_{\mathrm{ACT}} \in (0,1]$, critic learning rate $\lambda_{\mathrm{CRI}} \in (0,1]$, discount factor $\gamma \in (0,1]$, soft update coefficient $\tau \in (0,1]$, noise scale $\sigma$
\State Initialize actor network and critic network with random weights and both corresponding target networks.
\State Initialize exploration noise process $\mathcal{N}$
\For {episode $= 1, 2, \ldots$}
    \State Reset environment and obtain initial state $S^1$
    \State Reset noise process $\mathcal{N}$
    \For {$tt = 1, 2, \ldots, N_{\text{TTI}}$}
        \State Select action $A^{tt} = \mu(S^{tt}|\theta^{\mu}) + \mathcal{N}_\mathrm{DDPG}$
        \State Run $A^{tt}$, observe  $R^{tt+1}$ and $S^{tt+1}$
        \State Store $(S^{tt}, A^{tt}, R^{tt+1}, S^{tt+1})$ in replay buffer $\mathcal{M}$
        \If {$|\mathcal{M}| \geq N_{\text{b}}$}
            \State Sample minibatch $p$ of $N_{\text{b}}$ transitions from $\mathcal{M}$;
            \State Set $y^p = R^{p+1} + \gamma Q'(S^{p+1}, \mu'(S^{p+1}|\theta^{\mu'})|\theta^{Q'})$
            \State Update critic network;
            \State Update actor using the policy gradient;
            \State Update target networks;
        \EndIf
        \State $S^{tt} \leftarrow S^{tt+1}$
    \EndFor
\EndFor
\end{algorithmic}
\end{algorithm}

 \begin{figure*}[h!]
    \includegraphics[scale=0.65]{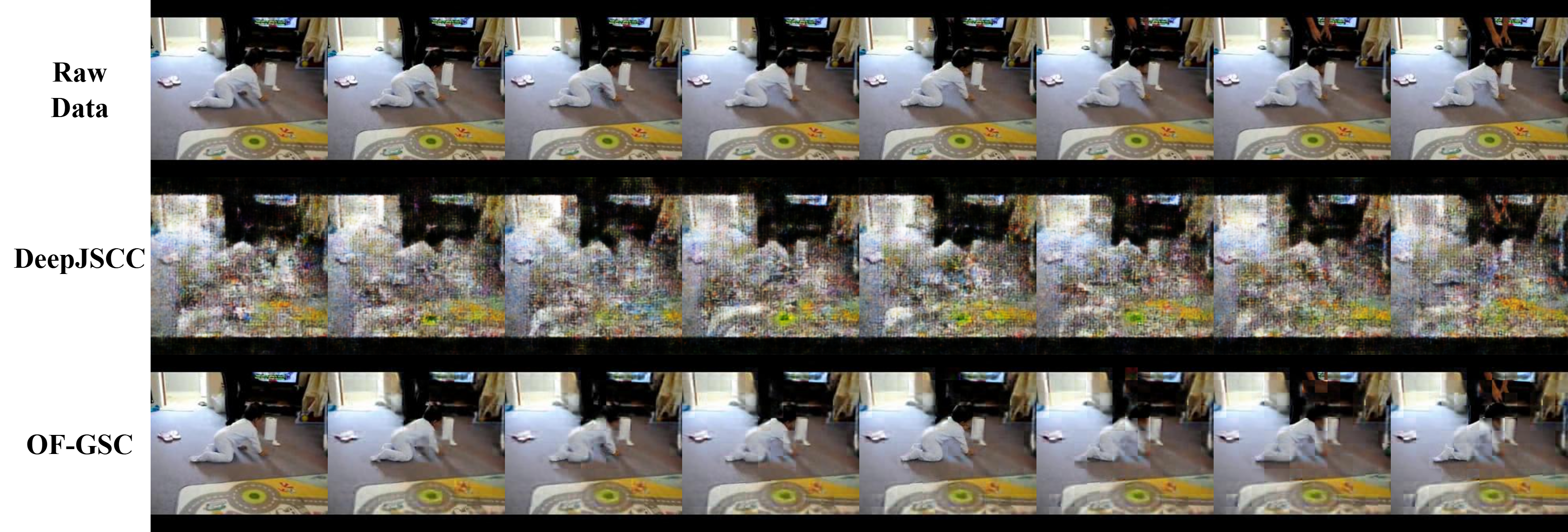}
    \centering
    \vspace{-0.1cm}
    \caption{Example of video reconstruction from DeepJSCC and OF-GSC for the same SNR.}
    \label{of_mae_demo}
    \vspace{-0.4cm}
\end{figure*}

\section{Performance Evaluation}

 In this section, we evaluate our proposed OF-GSC framework for wireless video reconstruction and video classification task through simulations. Our autoencoder model has been trained under different fixed mask ratios $\rho$, ranging from 0.0 to 0.9 with increments of 0.1. A random clip of 8 frames was selected from training videos to generate optical flow data. The training was performed under Python 3.10.11 and Pytorch 2.3.0 with four NVIDIA A100 40G GPUs \cite{hpc}. We employ the SSIM score to evaluate the video reconstruction quality and Top-1 Accuracy to evaluate the video classification accuracy.
 
\subsection{Training Dataset}
 During training, UCF-101\cite{ucf_101} and MPI Sintel\cite{sintal} datasets are implemented for simulation. UCF-101 contains 13,320 short ($<$10 second) trimmed videos from 101 action categories. Sintel is a dataset that contains all the frames from a 5-min video clip for optical flow network training. OF-GSC was trained for the video reconstruction task on both UCF-101 and Sintel, and for the classification task only on Sintel.
 \\

 \subsection{Reconstruction Quality of OF-GSC}

Fig. \ref{of_mae_demo} presents a frame-by-frame comparison between the original video frames and their reconstructions obtained by DeepJSCC and OF-GSC at an SNR of 0.78dB, with a mask ratio of 0.9 for OF-GSC. It is  observed that the frames reconstructed by DeepJSCC exhibit severe blurring and loss of fine details, which significantly reduces the visual quality and impairs motion perception under low SNR conditions. In contrast, the OF-GSC reconstructions demonstrate remarkable fidelity to the original frames, preserving both structural details and temporal consistency. This is because DeepJSCC maps each frame into a latent representation frame by frame at the bit-level, which can preserve a large amount of less important semantic representation in both spatial and temporal domain. However, OF-GSC has the ability to selectively reconstruct crucial semantic representation, particularly from motion-related patches, while omitting less informative regions. Consequently, it can be concluded that OF-GSC ensures superior video coherence and visual quality, even under challenging channel conditions, highlighting its advantage over conventional DeepJSCC in preserving both static and dynamic content.

 \begin{figure}[t]
\vspace{-0.3cm}
\centering
\setlength{\subfiglabelskip}{0pt}
\subfigure[\ SSIM vs. Mask Ratio (ideal channel) ]{
\includegraphics[scale=0.323]{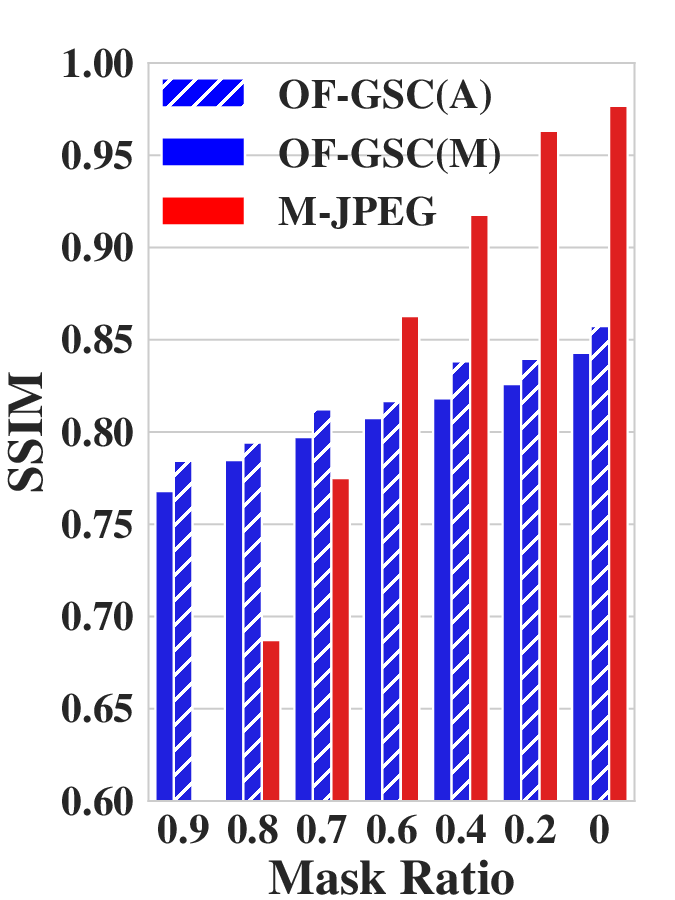}
\label{ssim_mr}
}
\subfigure[\ \centering SSIM vs. SNR]{
\includegraphics[scale=0.323]{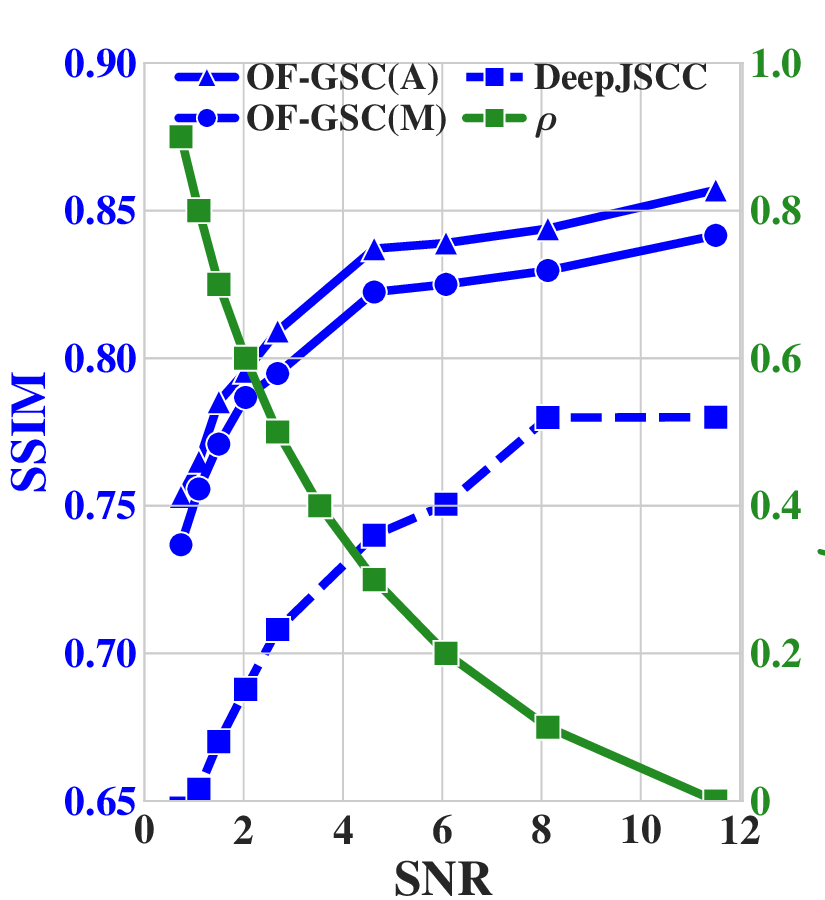}
\label{ssim_snr}
}

\subfigure[\ \centering PSNR vs. SNR]{
\includegraphics[scale=0.3]{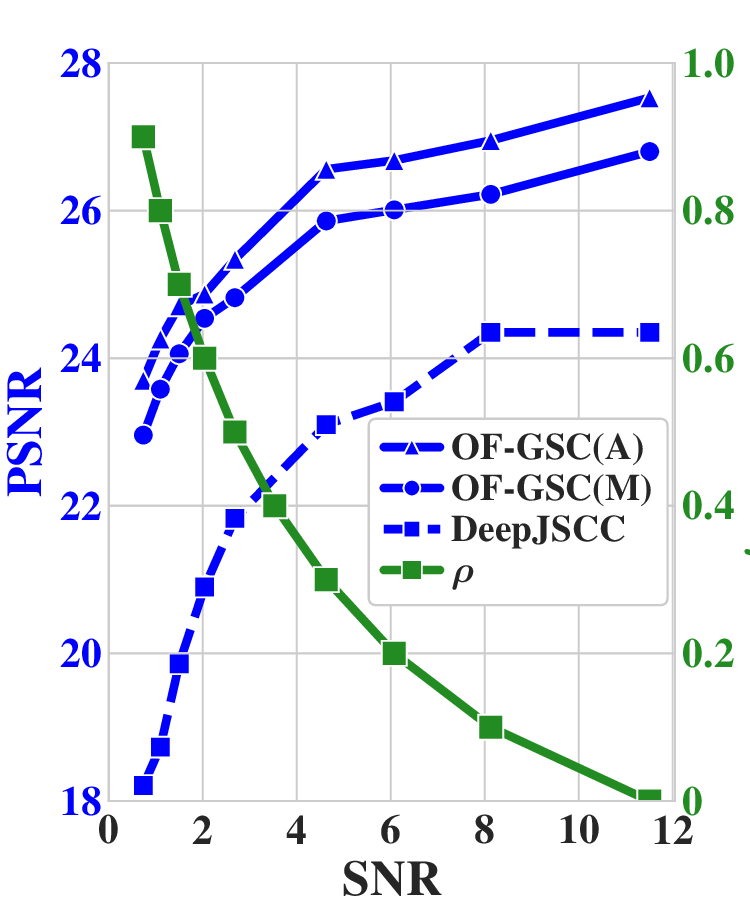}
\label{psnr_snr}
}
\subfigure[\ \centering VMAF vs. SNR]{
\includegraphics[scale=0.3]{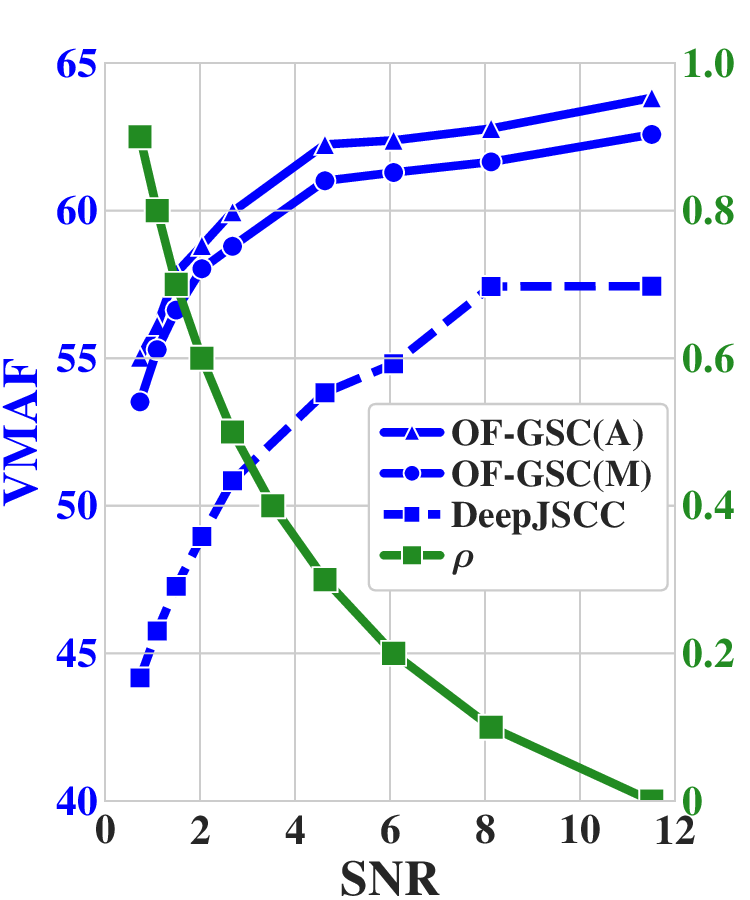}
\label{vmaf_snr}
}
\centering
\vspace{-0.2cm}
\caption{Metrics under various mask ratios and SNRs.}
\label{snr}
\vspace{-0.4cm}
\end{figure}

 In Fig. \ref{ssim_mr}, we compare the performance of OF-GSC and M-JPEG under the same transmission load and ideal wireless channels to validate the effectiveness of OF-GSC. The transmission load constraint for M-JPEG is determined by matching the total transmitted data volume of OF-GSC at different mask ratios. For OF-GSC, we test both OF-GSC with the proposed attention model (OF-GSC(A)) and OF-GSC using a mathematical motion detection algorithm (OF-GSC(M)). DeepJSCC is not taken into consideration in this simulation because its transmission load is fixed by its network structure. It can be observed that OF-GSC maintains higher stability and reconstruction quality under low SNR conditions, resulting in a 0.13 increase in SSIM at $\rho = 0.8$ compared with M-JPEG, while M-JPEG is unable to support the required transmission load when the mask ratio reaches $\rho = 0.9$. This performance difference is because M-JPEG relies on traditional bit transmission methods, implementing full video compression that inherently preserves substantial redundant information. Conversely, OF-GSC employs a selective transmission approach, targeting only the important semantic representation with minimal auxiliary information. This strategy significantly reduces the transmitted data volume, and makes OF-GSC more adaptable to constrained communication environments. It can also be observed that OF-GSC(A) demonstrates a modest performance gain over OF-GSC(M). This incremental improvement is because the mathematical motion detection semantic representation extractor follows fixed selection rules, which make it static and unable to capture patches with slight but meaningful motion, leading to the loss of content dependent details that are important for accurate reconstruction. In contrast, the attention based semantic representation extractor adapts its focus according to the actual video content and assigns higher importance to critical semantic representations, enabling more flexible and context aware reconstruction. Therefore, attention-based algorithm better helps OF-GSC to keep the important semantic representations for video reconstruction while discarding redundancy, which results in an advantage over the mathematical motion detection algorithm.

 \begin{figure}
    \vspace{-0.1cm}
    \includegraphics[scale=0.30]{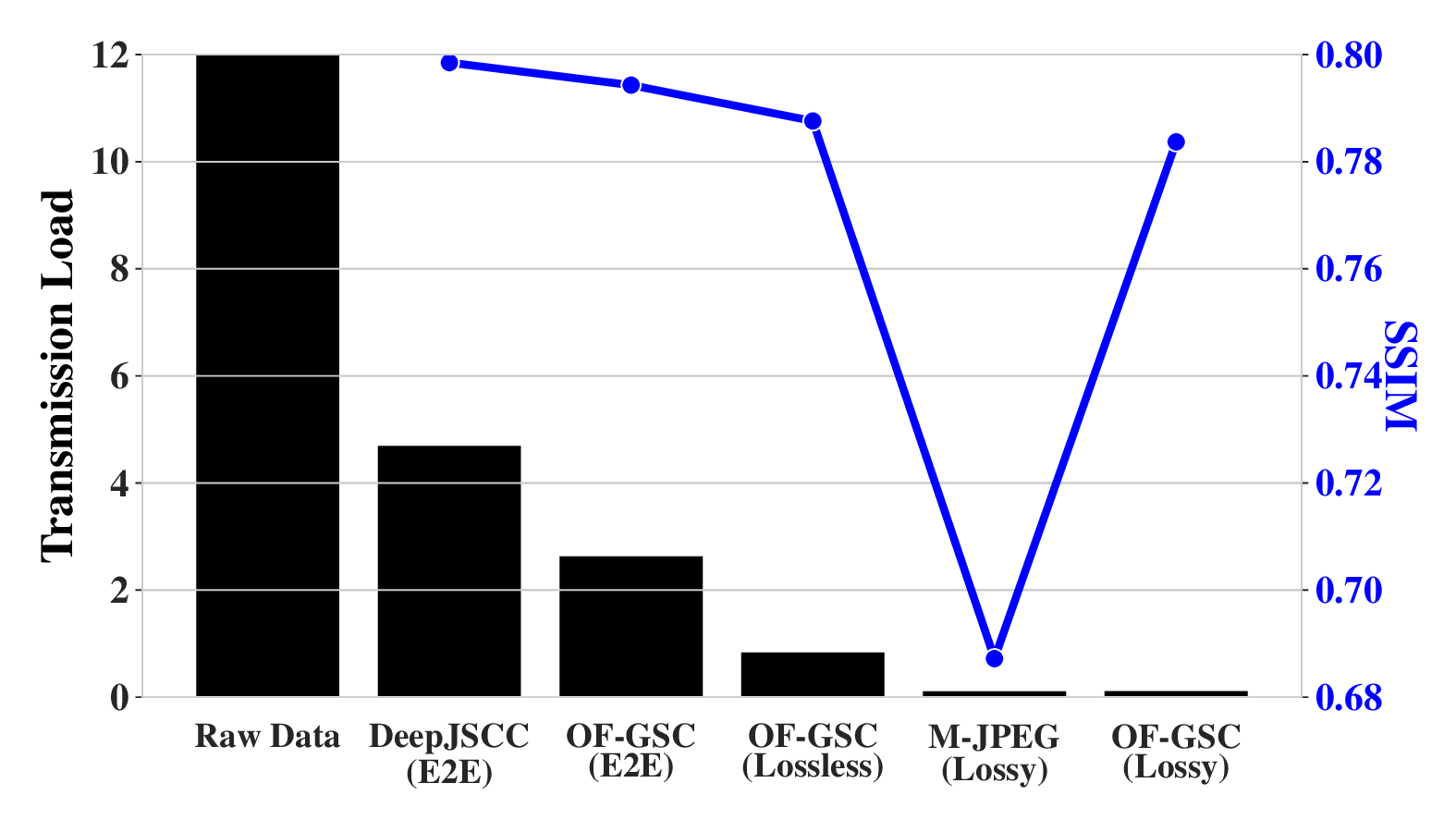}
    \centering
    \vspace{-0.08cm}
    \caption{Transmission load and SSIM score for ideal channels.}
    \label{compare_total}
\end{figure}

 In Fig. \ref{ssim_snr}, \ref{psnr_snr}, and \ref{vmaf_snr}, we compare the performance of OF-GSC and DeepJSCC under different SNR conditions. M-JPEG is not taken into consideration because it is highly susceptible to wireless channel errors. We select OF-GSC with different mask ratios based on the SNR. We can see that OF-GSC significantly outperforms DeepJSCC in reconstruction quality, particularly at SNR levels below 2dB, where OF-GSC improves the SSIM score by more than 0.1. This performance improvement is due to DeepJSCC's fundamental limitation in exploiting temporal frame correlations. OF-GSC, by contrast, effectively leverages both inter-frame and intra-frame information to compensate for missing data in the current frame, thereby achieving more robust reconstruction quality.

\subsection{Transmission Load Analysis}

Fig. \ref{compare_total} presents a comparative evaluation of the transmission load and corresponding SSIM scores for DeepJSCC, M-JPEG, and OF-GSC(A) ($\rho=0.8$) for ideal channel conditions. Three encoding schemes applied before transmission are considered, namely, end-to-end encoding (E2E), lossless encoding (Lossless), and lossy encoding (Lossy). We can observe that our E2E OF-GSC implementation achieves a 43.77\% reduction in data transmission volume compared to DeepJSCC with E2E, while maintaining comparable SSIM scores above 0.79.  This observed advantage is mainly because DeepJSCC encodes the raw image data. However, OF-GSC encodes the selected semantic representation patch, which provides a more compact representation and reduces spatial and temporal redundancy compared with raw image data. We can also see that lossy OF-GSC outperforms M-JPEG with a 0.1 improvement in terms of SSIM score, given the same transmission load constraint. This superior performance is because M-JPEG compresses video frames at the bit-level, which allows substantial redundancy to persist in entropy coding and results in a high transmission load. In contrast, the innovative optical flow-based motion information extractor and semantic representation extractor within OF-GSC effectively capture the semantically most relevant representation for video reconstruction, thereby reducing temporal and spatial redundancies and enabling highly efficient video quality preservation. Furthermore, it can be observed that lossless OF-GSC achieves 0.01 higher SSIM score compared with lossy OF-GSC, at the cost of a sixfold increase in transmission load. This is because lossless OF-GSC keeps all optical flow details, which results in a large amount of transmitted data, while lossy OF-GSC transmits optical flow that remains robust to slight distortions. Therefore, it can be concluded that lossy OF-GSC provides a more balanced and efficient choice for practical, transmission load constrained scenarios.

\subsection{Ablation Study}

\begin{figure}
    \includegraphics[scale=0.30]{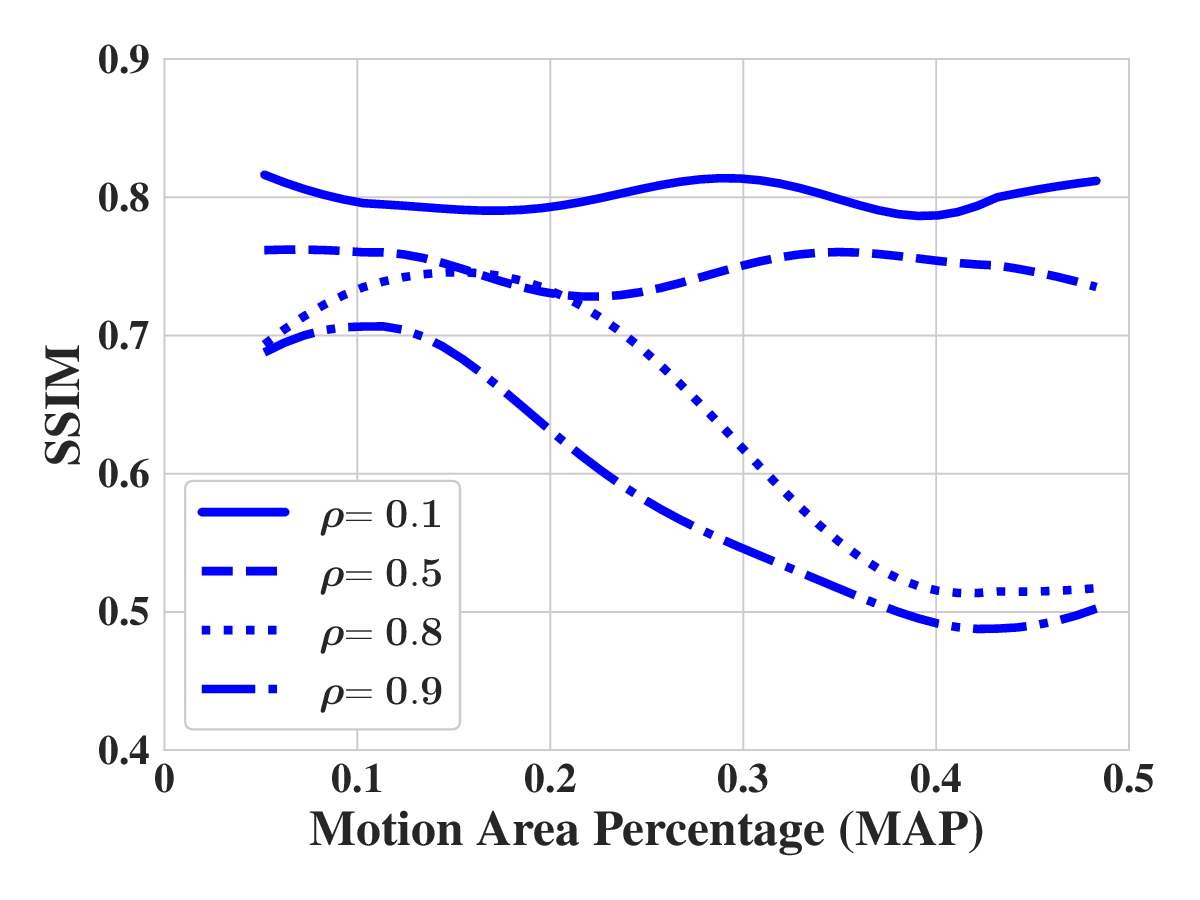}    \centering
    \vspace{-0.2cm}
    \caption{Impact of various motion area percentage.}
    \label{frame_gap}
    \vspace{-0.4cm}
\end{figure}

Fig. \ref{frame_gap} illustrates the effect of the motion area percentage (MAP) on the video reconstruction quality of OF-GSC frameworks with different mask ratios $\rho$, where MAP represents the proportion of significant semantic representation patches relative to the total number of patches. It can be seen that OF-GSC models with different mask ratios all achieve similar SSIM values,  when MAP is relatively low (about 5\%). This is because OF-GSC can robustly reconstruct videos when the amount of significant motion information is limited. However, we can observe that schemes with lower mask ratios obtain higher SSIM scores than those with higher mask ratios, and the gap reaches 0.25 at MAP equal to 0.4. This is because frameworks with higher mask ratios cannot cover enough significant  semantic representation patches for video reconstruction, while frameworks with lower mask ratios have enough significant semantic representation patches to maintain high reconstruction fidelity. Therefore, we can conclude that each mask ratio corresponds to a specific MAP threshold. When the MAP value stays below this threshold, the reconstruction quality remains high. Once it rises above the threshold, the SSIM score drops rapidly because the transmitted semantic representation becomes insufficient. This threshold behavior reflects the stability of the motion information contained in optical flow, which continues to support reliable reconstruction within the valid MAP range and thereby illustrates the suitability of optical flow in this framework.

\begin{figure}
    \includegraphics[scale=0.4]{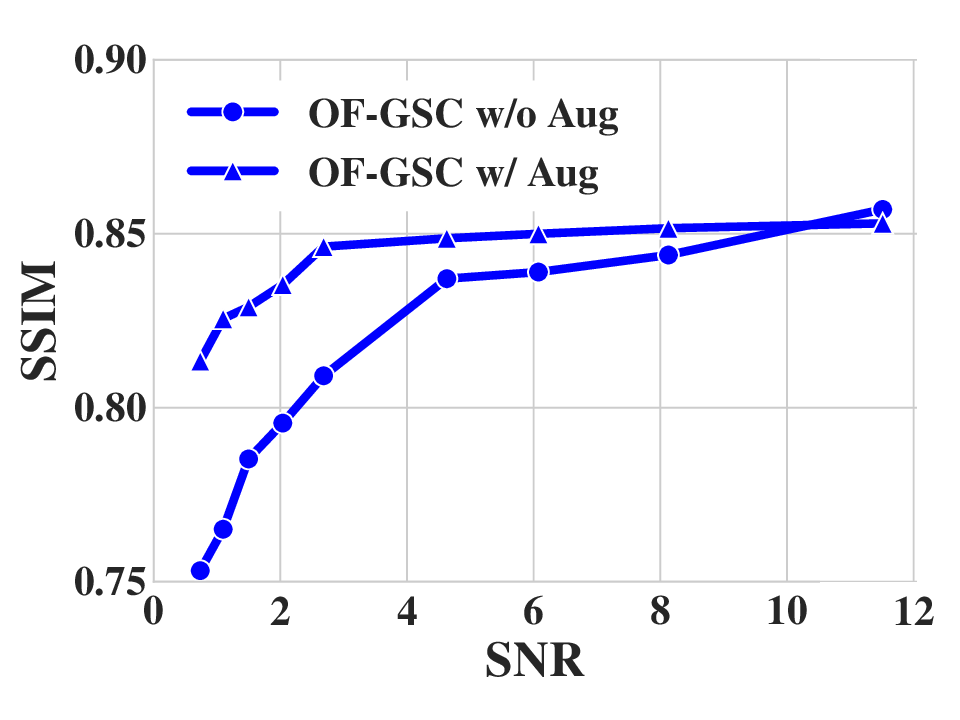}    \centering
    \vspace{-0.2cm}
    \caption{The SSIM score w/o Video Quality Augmentation model.}
    \label{aug}
    \vspace{-0.5cm}
\end{figure}

Fig. \ref{aug} illustrates the SSIM scores for video quality augmentation models. OF-GSC(A) with augmentation (OF-GSC w/ Aug) significantly outperforms OF-GSC(A) without augmentation (OF-GSC w/o Aug), with 0.07 higher SSIM score under low SNR conditions, demonstrating the effectiveness of the augmentation module in mitigating inter-patch boundary artifacts when limited information is transmitted.The improvement is because the augmentation module rebuilds the lost correlation across adjacent patches. The module focuses on the patch-level connection within the frame. Once this cross patch connection is recovered, artificial edges fade and the reconstructed frame becomes more coherent.

\subsection{Classification Quality of OF-GSC}

 \begin{figure}[t]
\centering
\setlength{\subfiglabelskip}{0pt}
\subfigure[Top-1 Accuracy ]{
\includegraphics[scale=0.35]{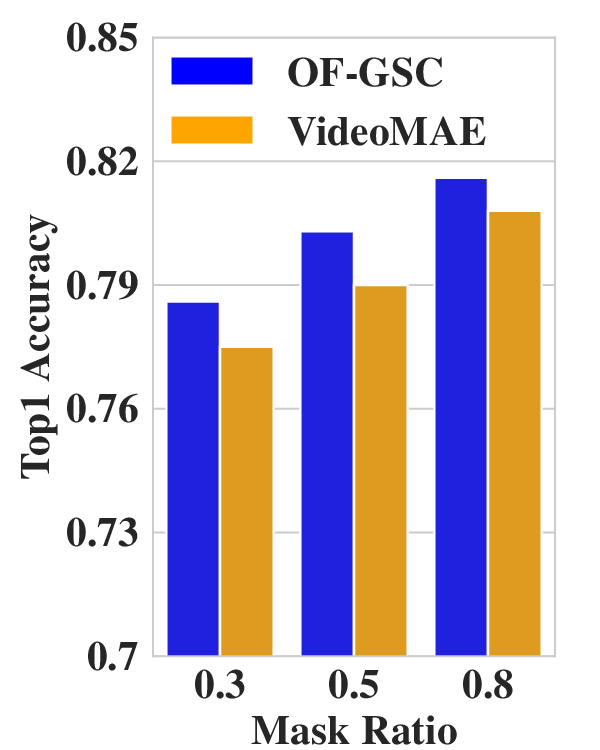}
\label{acc_res}
}
\subfigure[\ \centering Transmission Load]{
\includegraphics[scale=0.35]{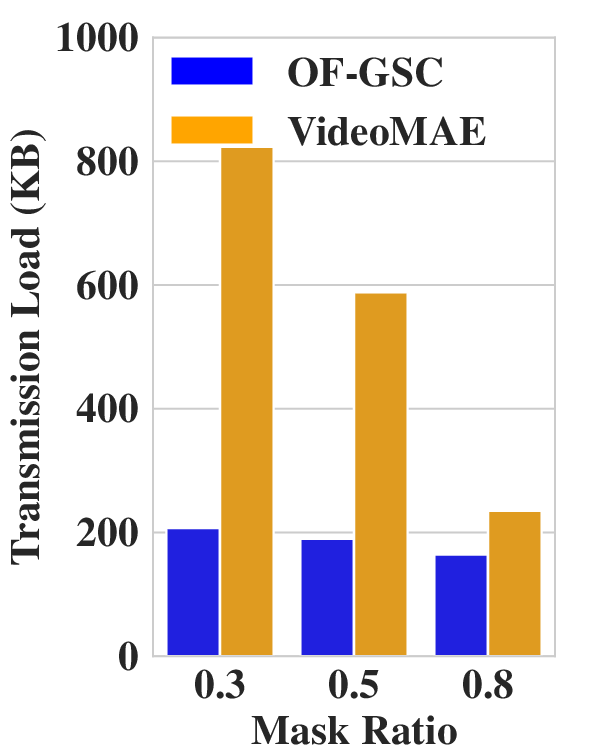}
\label{acc_load}
}
\centering
\vspace{-0.2cm}
\caption{Top-1 accuracy and required transmission load for classification task.}
\label{classification}
\vspace{-0.4cm}
\end{figure}

Fig. \ref{classification} presents the Top-1 classification accuracy and the required transmission load of OF GSC under different mask ratios compared with VideoMAE. It can be observed that OF-GSC attains a slightly higher Top-1 accuracy (about 1\%) while requiring less than 200KB of transmission load, which is far smaller than the load required by VideoMAE. This result is because VideoMAE masks patches at random and bases its classification on pixel patches, which still retain temporal redundancy. In contrast, OF-GSC selects the important semantic representations and relies on the first frame together with optical flow patches for classification. It can also be seen that the reduction in load for OF-GSC is less drastic because OF-GSC sends the full first frame as the base knowledge. Even so, the optical flow data remain much lighter than raw pixel data, which leads to a total transmission load that stays well below that of VideoMAE. This is because VideoMAE learns the spatial and temporal correlation among pixel patches. In contrast, OF-GSC obtains spatial correlations within the first frames and the important optical flow semantic representation provides temporal correlation across frames. Therefore, it is concluded that OF-GSC focuses on important semantic representation patches, which yields a compact feature set and supports higher classification accuracy while requiring a much lower transmission load.

\subsection{Bandwidth Allocation}
\begin{figure}[t]
\centering
\setlength{\subfiglabelskip}{0pt}
\subfigure[\ Bandwidth allocation ratios under different Scenarios]{
\includegraphics[scale=0.25]{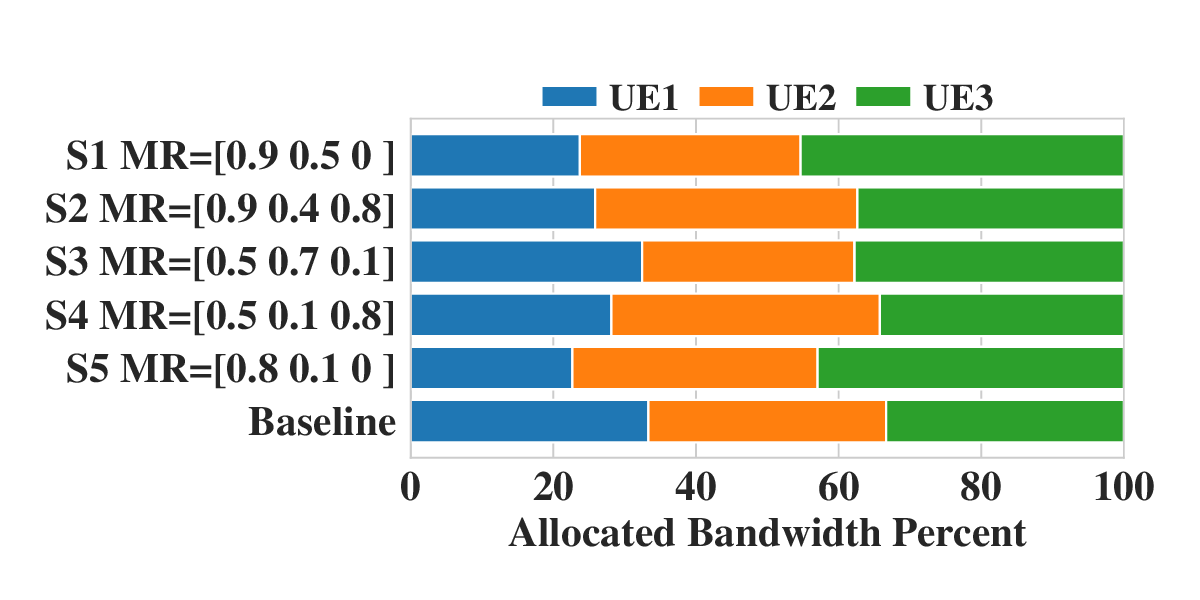}
\label{ba_pie}
}
\subfigure[\ \centering Maximum transmission time]{
\includegraphics[scale=0.25]{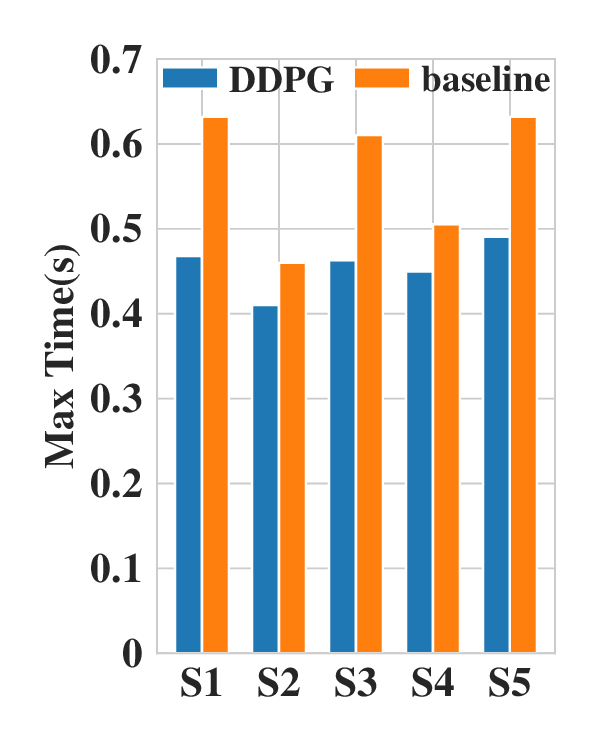}
\label{ba_bar}
}
\centering
\vspace{-0.2cm}
\caption{Bandwidth allocation simulation results.}
\label{ba}
\vspace{-0.4cm}
\end{figure}

Fig. \ref{ba} presents the maximum transmission time required by the DDPG-based bandwidth allocation algorithm and a baseline for scenarios with different mask ratios, where each UE maintains a fixed but different distance to the BS across all scenarios. The baseline adopts an average bandwidth allocation strategy for 3 UEs. Fig. \ref{ba_pie} presents the detailed allocation ratios, while Fig. \ref{ba_bar} shows the maximum transmission time of the proposed DDPG-based algorithm compared with the baseline across five different scenarios. It can be seen that our bandwidth allocation algorithm has a clear advantage over the baseline for most scenarios, and in Scenario 1, we achieve a 25.97\% reduction in maximum transmission time compared with the baseline scheme. Because the average strategy does not take the distance and data volume into consideration, the baseline does not consider the difference in distance or the number of transmitted bits, while our bandwidth allocation algorithm acquires distance information during training and adjusts the bandwidth according to the mask ratio. Therefore, as evidenced by Fig. \ref{ba_pie}, the DDPG-based bandwidth allocation algorithm assigns more resources to users with heavy transmission load and fewer resources to those with light load, which reduces waiting time and improves system-level efficiency.

\section{Conclusion}
\label{conc}

In this paper, we developed an efficient modular optical flow-based goal-oriented semantic communication framework for multi-user multi-task video transmission. To this end, an optical flow–based semantic encoder was designed to identify and transmit important semantic representation patches, together with a semantic decoder that supports multi-task video transmission. To minimize the maximum transmission time under multi-user conditions, a DDPG-based bandwidth allocation algorithm was developed. For video reconstruction tasks, we conducted experiments to compare our framework with DeepJSCC and M-JPEG. Compared with M-JPEG, our framework achieves an SSIM gain that reaches up to 14.04\% under stringent transmission load constraints and can reconstruct the video in conditions that M-JPEG is unable to support. Compared with DeepJSCC, our framework achieves a consistently higher SSIM score under different SNRs, with the largest gain reaching 0.1. For video classification tasks, our framework reaches a similar Top-1 Accuracy with a 25\% reduction in transmission load at a mask ratio of 0.3 compared with VideoMAE. For the bandwidth allocation problem, our DDPG-based algorithm yields a much smaller maximum transmission time compared with a baseline in most situations, with a reduction of about 25.97\% in one representative case. This work opens new avenues for research into integrating advanced masked autoencoder models for selection and transmission of efficient video semantic representation.


%





\ifCLASSOPTIONcaptionsoff
  \newpage
\fi



%
\bibliographystyle{IEEEtran}
\bibliography{mybib}

\begin{thebibliography}{99}
\BIBentryALTinterwordspacing
\bibitem{overall}
Cisco, Visual Network Index. "Cisco visual networking index: Forecast and methodology, 2016–2021." \emph{CISCO White paper}, 2022.
\bibitem{overall_1}
X.~Song, W.~Li, B.~Li, X.~Cheng, and C.~Dou,
"Optimization of Real-Time Data Transmission and Synchronization Mechanism in Audio and Video Conferencing Systems,"
\emph{2024 Int. Conf. Telecommun. Power Electron. (TELEPE)}, pp.~736--741,
Frankfurt, Germany, May~2024.
\bibitem{overall_1_1}
A.~Danesh~Pazho, et al.,
"Ancilia: Scalable Intelligent Video Surveillance for the Artificial Intelligence of Things,"
\emph{IEEE Internet Things J.}, vol.~10, no.~17, pp.~14940--14951, Sept.~2023.
\bibitem{overall_1_2}
A.~Alhilal, T.~Braud, B.~Han, and P.~Hui,
"Nebula: Reliable Low-latency Video Transmission for Mobile Cloud Gaming,"
\emph{Proc. ACM Web Conf. (WWW)}, pp.~3407--3417,
New York, NY, USA, 2022.
\bibitem{overall2}
Z. Wang, Y. Deng, and A. Hamid Aghvami, “Goal-oriented semantic communications for avatar-centric augmented reality,” \emph{IEEE Trans. Commun.}, pp. 1–1, Jun. 2024.
\bibitem{overall2_1}
H.~Xiao, et al.,
"A Transcoding-Enabled 360° VR Video Caching and Delivery Framework for Edge-Enhanced Next-Generation Wireless Networks,"
\emph{IEEE J. Sel. Areas Commun.}, vol.~40, no.~5, pp.~1615--1631, May~2022.
\bibitem{mjpeg}
ISO/IEC, "Information technology -- Digital compression and coding of continuous-tone still images -- Part 1: Requirements and guidelines," \emph{ISO/IEC 10918-1}, 1994.

\bibitem{mjpeg2000}
ISO/IEC, "Information technology -- JPEG 2000 image coding system -- Part 3: Motion JPEG 2000," \emph{ISO/IEC 15444-3}, 2002.
\bibitem{dirac}
T.~Borer and T.~Davies, "Dirac Video Compression Using Open Standards," \emph{BBC White Paper}, WHP~154, BBC Research \& Development, Jul.~2005.
\bibitem{h264}
ITU-T and ISO/IEC, "Information technology -- Coding of audio-visual objects -- Part 10: Advanced Video Coding," \emph{ITU-T Recommendation H.264 | ISO/IEC 14496-10}, 2003.

\bibitem{h265}
ITU-T and ISO/IEC, "Information technology -- High efficiency video coding," \emph{ITU-T Recommendation H.265 | ISO/IEC 23008-2}, 2013.

\bibitem{overall3}
Y.~Deng, Y.~Liu, N.~Pappas, J.~Zhang, Y.~Wang, and Y. Wang, “Guest editorial: Task-oriented communications and networking for the internet of things,” \emph{IEEE Internet Things Mag.}, vol. 6, no. 4, pp. 8–9, Dec. 2023.
\bibitem{overall4}
S.~Liu, N.~Li, Y.~Deng, and T.~Q.~S.~Quek,
"Goal-Oriented Semantic Communication for Wireless Visual Question Answering,"[Online], Nov.~2024
arXiv preprint arXiv:2411.02452.
\bibitem{overall5}
\bibitem{zhou2021uav}
H.~Zhou, F.~Hu, M.~Juras, A.~B.~Mehta, and Y.~Deng,
"Real-Time Video Streaming and Control of Cellular-Connected UAV System: Prototype and Performance Evaluation,"
\emph{IEEE Wireless Commun. Lett.}, vol.~10, no.~8, pp.~1657--1661, Aug.~2021.
\bibitem{sc_1}
B. Xu, R. Meng, Y. Chen, X. Xu, C. Dong, and H. Sun, “Latent semantic diffusion-based channel adaptive de-noising semcom for future 6g systems,” in \emph{Proc. IEEE Global Commun. Conf. (GLOBECOM)}, Kuala Lumpur, Malaysia, Dec. 2023, pp. 1229–1234.
\bibitem{sc_2}
X. Luo, H.-H. Chen, and Q. Guo, “Semantic communications: Overview, open issues, and future research directions,” \emph{IEEE Wireless Commun.}, vol. 29, no. 1, pp. 210–219, Jan. 2022.
\bibitem{sc_3}
H.~Zhou, Y.~Deng, X.~Liu,N.~Pappas,A.~Nallanathan,"Goal-Oriented Semantic Communications for 6G Networks", \emph{IEEE Internet Things Mag.}, vol. 7, no. 5, pp. 104-110, Sep. 2024.
\bibitem{sc_4}
M.~Kountouris and N.~Pappas,
"Semantics-empowered communication for networked intelligent systems,"
\emph{IEEE Commun. Mag.}, vol.~59, no.~6, pp.~96--102, Jun.~2021.

 \bibitem{sc_video_img}
 P. Jiang, C. K. Wen, S. Jin and G. Y. Li, "Wireless Semantic Communications for Video Conferencing,"  \emph{IEEE Journal on Selected Areas in Communications}, vol. 41, no. 1, pp. 230-244, Jan. 2023. 
\bibitem{deepjscc}
E.~Bourtsoulatze,D.~Burth Kurka,D.~Gündüz, "Deep Joint Source-Channel Coding for Wireless Image Transmission", \emph{IEEE Trans. Cogn. Commun. Netw.}, vol. 5, no. 3, pp. 567-579, Sept. 2019.
\bibitem{deepjscc_lm}
Y. Liu, Z. Qin, S. Wang, C. Wang, "Dynamic Feature Transmission for Image Communication via Deep Joint Source-Channel Coding", \emph{Proceedings of the 2024 14th International Conference on Communication and Network Security
(ICCNS24)}, pp. 139-145, Feb. 2025.
\bibitem{adjscc}
J.~Xu, B.~Ai, W.~Chen, A.~Yang, P.~Sun, and M.~Rodrigues, "Wireless image transmission using deep source channel coding with attention modules", \emph{IEEE Trans. Circuits Syst. Video Technol.},  vol. 32, no. 4, pp. 2315-2328, April 2022.
\bibitem{attn_video}
N.~Li, M.~Bennis, A.~Iosifidis, and Q.~Zhang,"Spatiotemporal attention-based semantic compression for real-time video recognition", \emph{Proc. IEEE Global Commun. Conf. Workshops(GC Wkshps)}, Kuala Lumpur, Malaysia, Dec. 2023, pp. 1603-1608.

\bibitem{temporal_video_2}
Z. Zhang, Q. Yang, S. He and J. Chen, "Deep Learning Enabled Semantic Communication Systems for Video Transmission," \emph{2023 IEEE 98th Veh. Technol. Conf. (VTC2023-Fall)}, Hong Kong, 2023, pp. 1-5
\bibitem{gop}
B. Zatt, M. Porto, J. Scharcanski and S. Bampi, "Gop structure adaptive to the video content for efficient H.264/AVC encoding," \emph{2010 IEEE International Conference on Image Processing}, Hong Kong, China, 2010, pp. 3053-3056
\bibitem{gsc_1}
E.~C.~Strinati, P.~Di~Lorenzo, V.~Sciancalepore, A.~Aijaz, M.~Kountouris, D.~Gündüz, P.~Popovski, M.~Sana, P.~A.~Stavrou, B.~Soret, et al.,
"Goal-oriented and semantic communication in 6G AI-native networks: The 6G-GOALS approach,"
\emph{arXiv preprint arXiv:2402.07573}, Feb.~2024.

\bibitem{gsc_2}
E.~C.~Strinati and S.~Barbarossa,
"6G networks: Beyond Shannon towards semantic and goal-oriented communications,"
\emph{Comput. Netw.}, vol.~190, p.~107930, 2021.
\bibitem{gsc_3}
Z.~Wang and J.~Zhao,
"Utility-Oriented Communications for 6G Mobile Networks and the Metaverse: Semantic, Task-Oriented, Goal-Oriented, and More,"
\emph{2023 IEEE 43rd Int. Conf. Distrib. Comput. Syst. (ICDCS)}, pp.~1--2,
Hong Kong, Hong Kong, 2023.
\bibitem{gsc_4}
F.~Z.~Safaeipour and M.~Hashemi,
"Semantic-Aware and Goal-Oriented Communications for Object Detection in Wireless End-to-End Image Transmission,"
\emph{2024 Int. Conf. Comput., Netw. Commun. (ICNC)}, pp.~182--187,
Big Island, HI, USA, 2024.
\bibitem{spynet}
A.~Ranjan, M.~J.~Black. "Optical Flow Estimation Using a Spatial Pytamid Network", \emph{Proc. IEEE Conf. Comput. Vis. Pattern Recognit.(CVPR)}, Honolulu, Hawaii, July 2017, pp. 4161-4170.
\bibitem{of_algo}
J. Huang, W. Zou, Z. Zhu and J. Zhu, "An Efficient Optical Flow Based Motion Detection Method for Non-stationary Scenes". \emph{2019 Chinese Control Decision Conf. (CCDC)}, Nanchang, China, June 2019, pp. 5272-5277.
\bibitem{pos_embed}
Y.~Huang, Z.~Zhao, H.~H, etc. ."Attention is all you need: A Pytorch Implementation". Availible: https://github.com/jadore801120/attention-is-all-you-need-pytorch/blob/master/transformer/Models.py

\bibitem{hpc}
King's~College~London, "King's Computational Research, Engineering and Technology Environment (CREATE)". March 2, 2022, Availible:  https://doi.org/10.18742/rnvf-m076

\bibitem{plus1}
N.~Li, A.~losifidis, Q.~Zhang, "Dynamic Semantic Compression for CNN Inference in Multi-access Edge Computing: A Graph Reinforcement Learning-based Autoencoder". \emph{IEEE Trans. Wirel. Commun.}, Dec. 2024.

\bibitem{plus2}
N. Li, M. Bennis, A. Iosifidis and Q. Zhang, "Spatiotemporal Attention-based Semantic Compression for Real-time Video Recognition,"\emph{2023 IEEE Globecom Workshops (GC Wkshps)}, Kuala Lumpur, Malaysia, 2023, pp. 1603-1608.
\end{thebibliography}
%








\end{document}